\documentclass[11pt,a4paper,onecolumn,reqno]{amsart}
\usepackage[a4paper, margin=2.3cm, bmargin=3cm]{geometry}

\usepackage{graphicx}
\usepackage{dcolumn}
\usepackage{bm}

\usepackage[utf8]{inputenc}
\usepackage[T1]{fontenc}
\usepackage{mathptmx}
\usepackage{amsaddr}
\usepackage[colorinlistoftodos,prependcaption,textsize=tiny]{todonotes}
\usepackage{gensymb}
\usepackage{threeparttable}
\usepackage{color, soul} 
\usepackage{caption}
\usepackage{subcaption}
\usepackage{verbatim}
\usepackage{graphicx,color,psfrag,amssymb,amsmath,url}
\usepackage{bm}
\usepackage{booktabs}
\usepackage{tabularx}
\usepackage{multirow}
\usepackage{xcolor}
\usepackage{hyperref}
\usepackage{placeins}
\usepackage[version=4]{mhchem} 

\usepackage[square,comma,sort&compress, numbers]{natbib}

\usepackage{caption}
\usepackage{subcaption}
 \usepackage{pict2e}
\newcommand{\dpartial}[2]{\frac{\partial #1}{\partial #2}}

\newcommand{\colorblue}{\color{black}}
\newcommand{\colorblack}{\color{black}}
\usepackage{etoolbox}
\patchcmd{\pprintMaketitle}
  {\hrule\vskip12pt}
 {\hrule\vskip12pt\ifvoid\extrainfobox\else\unvbox\extrainfobox\par\vskip12pt\fi}
 {}{}

\newsavebox\extrainfobox

\title{Soot particle size distribution reconstruction in a turbulent sooting flame with the  split-based extended Quadrature method of moments}
\author[]{Federica Ferraro$^{1,*}$,  Sandro Gierth$^1$,  Steffen Salenbauch,Wang Han$^2$, Christian Hasse$^1$ }
\email{ferraro@stfs.tu-darmstadt.de} 
\address[]{$^1$Technical University of Darmstadt, Department of Mechanical Engineering, Simulation of reactive Thermo-Fluid Systems, Otto-Berndt-Straße 2, Darmstadt 64287, Germany \\
$^2$School of Engineering, University of Edinburgh, Edinburgh EH8 3JL, Scotland, UK}

\begin{document}

\pagestyle{plain}

\begin{large}

\begin{figure}[t]
\centering
\includegraphics[scale=0.9]{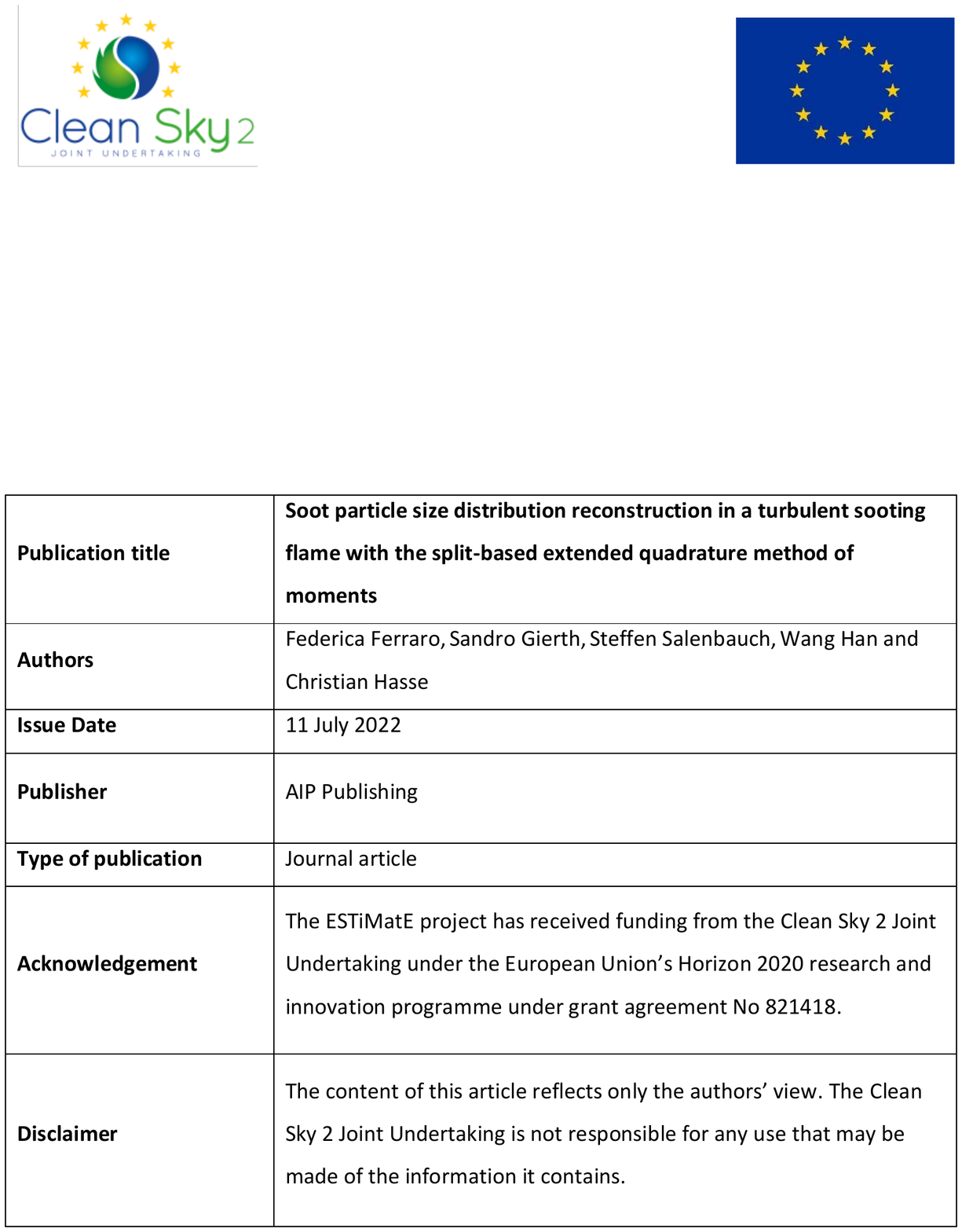}
\caption*{}
\end{figure}
\FloatBarrier
\vspace*{8cm}
This article may be downloaded for personal use only. Any other use requires prior permission of the author and AIP Publishing. This article appeared in 

\noindent \textbf{Federica Ferraro, Sandro Gierth, Steffen Salenbauch, Wang Han, and Christian Hasse , "Soot particle size distribution reconstruction in a turbulent sooting flame with the split-based extended quadrature method of moments", Physics of Fluids 34, 075121 (2022)}  

\noindent and may be found at 
 \href{https://doi.org/10.1063/5.0098382}{\textbf{https://doi.org/10.1063/5.0098382}}. 
 
\end{large}
 \newpage

\maketitle

\begin{abstract}
The Method of Moments (MOM) has largely been applied to investigate sooting laminar and turbulent flames. 
However, the classical MOM is not able to characterize a continuous particle size distribution (PSD). 
Without access to information on the PSD, it is difficult to accurately take into account particle oxidation, which is crucial for shrinking and eliminating soot particles. Recently, the Split-based Extended Quadrature Method of Moments (S-EQMOM) has been proposed as  a numerically robust alternative to overcome this issue (Salenbauch et al., 2019).  The main advantage is that a continuous particle number density function can be reconstructed by superimposing kernel density functions (KDF). Moreover, the S-EQMOM primary nodes are determined individually for each KDF, improving the moment realizability. 

In this work, the S-EQMOM is combined with a Large Eddy Simulation/presumed-PDF flamelet/progress variable approach for predicting soot formation in the Delft Adelaide Flame III. The target flame features low/high sooting propensity/intermittency and  comprehensive flow/scalar/soot data are available for model validation. 
Simulation results are compared with the experimental data for both the gas phase and the particulate phase. A good quantitative agreement has been obtained especially in terms of the soot volume fraction. The reconstructed PSD reveals predominantly unimodal/bimodal distributions in the first/downstream portion of  this flame, with  particle diameters smaller than 100 nm. By investigating the instantaneous and statistical sooting behavior at the flame tip, it has been found that the experimentally observed  soot intermittency  is  linked to mixture fraction fluctuations around its stoichiometric value that exhibit a bimodal probability density function.
\end{abstract}

\maketitle

\section{Introduction}

Soot  formation is an undesired phenomenon in  the combustion  of hydrocarbon fuels. 
Its harmful effects on climate change and human health call for strict regulations in terms of both the soot volume fraction and the particle size distributions emitted by engines and combustors. 
Accurately predicting the particle formation processes and particle size distribution (PSD) is  therefore  fundamental when designing sustainable combustion devices such as gas turbines or internal combustion engines with reduced levels of pollutants. 
The strong coupling between the turbulent flow field, gas-phase chemical reactions especially involving soot precursors, and the solid phase yields a highly complex physico-chemical problem. 
The particle evolution is typically obtained by solving the population balance equation (PBE) that describes the spatial and temporal transport of the particle number density function (NDF). 
Due to the high dimensionality of the PBE, sectional methods (SMs) \cite{Grosschmidt2007,Netzell2007,Rodrigues2018, Grader2018,Tian2021,Cifuentes2020} and the Method of Moments (MOM)~\cite{Zucca2006,Attili2014, Mueller2012a, Xuan2015,Koo2017} are the most widely used  approaches  to characterize the particle  NDF in turbulent reacting flows. 
Specifically, sectional methods  are based on the discretization of the particle size distribution in several sections representing particles in a defined interval of the particle property space, e.g., particle volume. 
One or more scalars are transported to represent the particles in one section, yielding a set of transport equations. Therefore, the SM naturally provides local information on the NDF.  
On the other hand, the method of moments does not solve the NDF directly, but provides an approximation of the NDF solving only transport equations for its low-order moments. This yields a computationally efficient strategy to evaluate the particle population that is suitable to be combined in reacting Large Eddy Simulations (LESs) and Direct Numerical simulations (DNSs). 
\colorblue
Other approaches have been  recently developed that are able to predict the  particle size distribution. Examples are an  LES filtered  density function  approach  extended by including the particle property distribution~\cite{Sewerin2017,Sewerin2018} and an hybrid stochastic/fixed-sectional method  proposed to solve the population balance equation~\cite{Seltz2021,Bouaniche2019a}. 
\colorblack

In recent years,  LESs~\cite{Mueller2012a,Han2018,Rodrigues2018,Sewerin2018,Tian2021} and DNSs \cite{Yoo2007,Lignell2007,Bisetti2012,Attili2014,Attili2016} of turbulent sooting  flames have improved soot prediction, yielding a deeper understanding of the interaction between turbulent reacting flows and soot formation. Reviews on numerical studies can be found in \cite{Raman2016a,Valencia2021,Rigopoulos2019}.
In~\cite{Bisetti2012,Attili2014}, a DNS of a turbulent non-premixed n-heptane/air  flame  was performed  to analyze the interactions of turbulent mixing, detailed chemical reactions, and soot formation. 
 The  Hybrid Method of Moments (HMOM)~\cite{Mueller2009a} was applied to evaluate the soot particle population.  Soot precursors were observed to be extremely sensitive to the local scalar dissipation rate, which  correlates to soot particle formation and  growth. 
In~\cite{Attili2016} it was found that when the differential diffusion of gas-phase species is neglected,  the soot number density is not altered significantly, while the soot mass yield decreases by a factor of two.
In \cite{Yoo2007}, non-premixed counterflow flames were calculated and the effect of the turbulence-chemistry interaction on the soot evolution was studied. 
It was found that  turbulence can raise the soot volume  fraction by increasing the flame volume, or reduce it by transporting the soot pockets out of the high-temperature regions. 
Several studies focus on  LESs of turbulent flames exploring different soot models, combustion and turbulence-combustion closures,  e.g., \cite{Mueller2012a,Donde2012a,Han2018,Rodrigues2018,Tian2021,Sewerin2017}.  However, uncertainties remain, related to the kinetics, particle dynamics, and their interaction with the turbulence~\cite{Rigopoulos2019}.
Few works have numerically investigated soot formation  in complex configurations such as  gas turbines~\cite{Mueller2013,Teng2018, Koo2017,Franzelli2018,Cokuslu2022,Tardelli2021,Eigentler2022,Wick2017b}.


Quadrature-based Methods of Moments (QbMOM) represent an alternative closure for moment transport equations~ \cite{Yuan2012,Salenbauch2015a,Salenbauch2016,Wick2017a,Ferraro2021a}. Here,  the unknown soot particle NDF is approximated either by a set of Dirac delta functions as in the Quadrature Method of Moments (QMOM)~\cite{McGraw1997} or by kernel density functions (KDFs)
as in the Extended Quadrature Method of Moments (EQMOM)~\cite{Chalons2010,Yuan2012}.
Of the different quadrature-based moment closures, the EQMOM approach~\cite{Yuan2012} has proven particularly suitable for the prediction of  soot  particle formation, growth, and oxidation~\cite{Wick2017a}, since it provides a continuous reconstruction of the NDF. Pointwise information on the NDF is indeed necessary to describe  the complete oxidation of the smallest soot particles~\cite{Wick2017a,Salenbauch2019}.   
However, the EQMOM algorithm presented some numerical issues~\cite{Pigou2018,Nguyen2016} that make the approach hard to integrate into  complex three-dimensional reacting simulations. 

Recently, following on from the idea proposed by Megaridis and Dobbins~\cite{Megaridis1990a} of considering the NDF as a sum of KDFs  interacting via the coagulation term, Salenbauch et al.~\cite{Salenbauch2019} 
have shown that the  numerical difficulties of the EQMOM can be overcome.  
The  so-called Split-based Extended QMOM approach (S-EQMOM)  represents an alternative Extended QMOM, where transport equations are only solved for the lower order moments of coupled KDFs (or sub-NDFs as in~\cite{Salenbauch2019}).
While in the EQMOM, an iterative and non-unique procedure~\cite{Nguyen2016, Pigou2018,Yuan2012}  is applied to invert the moments, from low and high order moments of the entire NDF, in S-EQMOM one-node EQMOM systems are solved for each sub-NDF yielding  a unique solution~\cite{Salenbauch2019}. 
Therefore, the S-EQMOM provides  a numerically robust algorithm to reconstruct the soot particle NDF. The validity of the method has been demonstrated in~\cite{Salenbauch2019} by simulating laminar premixed flames with unimodal and bimodal soot particle distributions and a two-stage burner flame configuration ~\cite{Echavarria2011a} under highly soot oxidizing conditions. 

Nevertheless, the performance of S-EQMOM in modeling soot formation and PSD in turbulent flames has not been examined.
With this background, the objective of this work is threefold: (i) to extend the S-EQMOM approach in the context of LES; (ii) to  examine the reconstructed particle size distribution by the S-EQMOM in a turbulent flame for the first time; and (iii) to analyze the dynamics of soot formation, growth and oxidation using the detailed results from the LES. 

The feasibility and accuracy of the S-EQMOM in turbulent sooting flames is first proven, then its capability to reconstruct the soot PSD is exploited to gain further insights into the physical processes affecting  the soot evolution.  

Here, the S-EQMOM  is combined with a tabulated flamelet/progress variable (FPV) approach and a presumed Beta-PDF, which provides the turbulence-chemistry interaction closure. 
The developed numerical framework is applied to simulate a turbulent natural gas/air flame, the Delft-Adelaide Flame III, which is one of the target flames of the ‘International  Sooting Flame Workshop’ (ISF)~\cite{Peeters1994,Qamar2009}. This flame exhibits  low sooting propensity with high intermittency, therefore it represents a challenging benchmark  for the soot model. 
The gas-phase flow field and soot properties are discussed and compared with the experimental data~\cite{Stroomer1995,Peeters1994,Qamar2009}. Subsequently, the PSD  is reconstructed and analyzed at different positions along the flame.  Finally, the soot dynamics is examined with respect to the mixture fraction field.



The extension of the S-EQMOM approach to the LES and the PSD reconstruction in turbulent flames constitute one of the  novelties of this work in the context of soot modeling. Furthermore, by exploiting the detailed information from the LES, special emphasis is given to the investigation of the  soot intermittency, which has a strong impact on the soot statistics in this slightly sooting flame.
In particular, it is found that the bimodal probability density function (PDF) of the maximum soot volume fraction at the flame tip, observed in the experiments~\cite{Qamar2009}, is linked to  a bimodal distribution of the mixture fraction~ PDF.

\section{Numerical Model}
\subsection{Method of Moments}
The evolution of the soot  NDF $n(t,\bm{x};\bm{\xi})$ is governed by the  population balance equation (PBE)
\begin{align}
\label{eq:PBE}
&\dpartial{n(t,\bm{x};\bm{\xi})}{t}
+ \dpartial{\bm{u}n(t,\bm{x};\bm{\xi})}{\bm{x}}
-\dpartial{}{\bm{x}} \left( 0.55 \dfrac{\nu}{T} \dpartial{T}{\bm{x}} n(t,\bm{x};\bm{\xi}) \right)
-\dpartial{}{\bm{x}}  \left(\Gamma_m \dpartial{n(t,\bm{x};\bm{\xi})}{\bm{x}} \right)
\\ & =  \nonumber
R_{nuc}+
R_{cond}+
R_{sg}+
R_{coag}+
R_{ox},
\end{align}
with $\bm{u}$ the velocity vector, $\nu$ the kinematic viscosity, $T$ the temperature, $\Gamma_m$ the Brownian diffusivity and $\bm{\xi}$ the vector of the internal coordinates. The source terms $R_Y$ with $Y \in [nuc, \, cond, \, sg, \, ox, \, coag]$ take into account the physical and chemical process of particle nucleation $R_{nuc}$, PAH condensation $R_{cond}$,  surface growth $R_{sg}$ by the HACA mechanism, particle coagulation $R_{coag}$ and oxidation $R_{ox}$. 
The fourth term on the left-hand side  represents 
the molecular diffusion transport, which is negligible due to the high Schmidt number of soot particles,  as shown by~\cite{Bisetti2012}, and is therefore not taken into consideration in this study.   
The vector of the internal coordinates  $\bm{\xi}$ contains the properties to characterize the soot particles. In this work, soot particles are considered to be spherical and are characterized only by their volume $V$,  $\bm{\xi}=[V]$. 

Due to the high dimensionality of the PBE, it is not feasible to solve it directly. Here, the Method of Moments is used, which only solves the evolution of low-order moments of the NDF.
The moment of order $k$ of the univariate NDF is defined as
\begin{equation}
m_k=\int_0^\infty V^k n(t,\bm{x};V) \mathrm{d}V.
\label{eq:moment def}
\end{equation}  
Applying  Eq. (\ref{eq:moment def}) to Eq.~(\ref{eq:PBE}), the transport equation for the $k-$th moment is obtained
\begin{equation}
\dpartial{m_k}{t}+
\dpartial{\bm{u}m_k}{\bm{x}}
-\dpartial{}{\bm{x}} \left( 0.55 \dfrac{\nu}{T} \dpartial{T}{\bm{x}} m_k \right)
=\dot{m}_k.
\end{equation}
The source terms $\dot{m}_k$ contain the  corresponding contributions of the  physico-chemical processes of particle formation, condensation, surface growth, coagulation and oxidation.  
Closures are required to determine the moment source terms, which depend on the unknown NDF. 
Here, the  Split-based Extended Quadrature Method of Moments (S-EQMOM) formulated in~\cite{Salenbauch2019} is employed to achieve the closure. 

  
\subsection{Split-based Extended QMOM}
In the standard QMOM method~\cite{MarchisioBook2013}, the unknown NDF is  approximated  by a linear combination of $N_V$ Dirac delta functions 
\begin{equation}
n(V)\approx \sum_{i=1}^{N_V}  w_i \delta (V-V_i),
\end{equation} 
where $w_i$ and $V_i$ are the Gaussian quadrature weights and nodes, respectively. 
The  determination of these parameters,  the so-called inversion of the moments, is  performed using suitable algorithms available in the literature,  such as the Wheeler algorithm~\cite{Wheeler1974}. 

In the EQMOM approach \cite{Yuan2012, MarchisioBook2013},  the NDF is approximated by a weighted sum of overlaying  kernel density functions $\delta_\sigma$, instead of delta functions,
\begin{equation}
n(\hat{V})\approx \sum_{i=1}^{N_{kdf}}  w_i \delta_{\sigma} \left(\hat{V} ;\hat{V}_i \right),
\end{equation}
where the  parameter $\sigma$, representing  the scale of the KDFs, additionally needs to be determined. 
$\delta_{\sigma}(\hat{V} ;\hat{V}_i )$ are the $N_{kdf}$ continuous KDFs, positioned at $\hat{V}_i $ and weighted by $w_i$, and are approximated by  kernel density functions of a known shape, e.g., gamma or lognormal distributions. 
Gamma distributions are commonly used to approximate the soot particle NDF~\cite{Wick2017a,Salenbauch2019}.  
A change of variable 
\begin{equation}
\hat{V}=V-V_{min}, \quad \hat{V}\in [0, \infty),
\end{equation}
is applied,  since the physical soot distribution is defined in the interval $V\in [V_{min},\infty)$,  with $V_{min}$ denoting the volume of a newly nucleated particle. 
\colorblue
The nodes $\hat{V}_i$, the weights $w_i$, and the shape parameter $\sigma$ are determined from $2N_{kdf}+1$ moments  using the moment-inversion algorithm  proposed by Yuan et al. \cite{Yuan2012}. 
\colorblack

The  S-EQMOM used in this work  is a recently developed alternative formulation of the EQMOM approach~\cite{Salenbauch2019} that,  while allowing a continuous PSD reconstruction similarly to the EQMOM,  leads to a numerically robust moment inversion procedure. The S-EQMOM indeed avoids the problems related to uniqueness and realizability 
\colorblue
discussed in \cite{Nguyen2016,Pigou2018}
 by solving for multiple coupled sub-NDFs rather than for the entire NDF and was validated  in laminar flames  in~\cite{Salenbauch2019}. 
This approach allows to take only  the  lower-order moments  $m_k^{s_i}$ of each  $N_s$ sub-NDFs $n_{s_i}(V)$  into consideration  instead of the higher order  moment of the entire NDF, $n(V)$.  
 \colorblack
A generic sub-NDF moment is defined as
\begin{equation}
m_k^{s_i}= \int_{V_{min}}^\infty V^k  n_{s_i}(V)\mathrm{d}V,
\end{equation}
where $s_i$ is the index  of the sub-NDF. 
The entire NDF may then be approximated as 
\begin{equation}
n(\hat{V})=\sum_{i=1}^{N_s}n_{s_i}(\hat{V}) \approx
\sum_{i=1}^{N_s} w_{s_i} \delta_{\sigma_{s_i}}(\hat{V}; \hat{V}_{s_i}) .
\end{equation}

\noindent In the S-EQMOM approach, the moment inversion procedure consists in analytically calculating  the value of the nodes   $V_{s_i}$, the weights $w_{s_i}$, and the scale parameters ${\sigma_{s_i}}$ directly from the  first three moments $[m_0^{s_i}, m_1^{s_i}, m_2^{s_i}]^T$ of the $N_s$ sub-NDFs. 
Note that this corresponds to the inversion of a series of one-node EQMOM systems 
\colorblue($N_{kdf}=$ 1) \colorblack
 using the solution algorithm proposed by Yuan et al.~\cite{Yuan2012}. The main advantage of the S-EQMOM over the EQMOM  is that the inversion procedure yields a system of equations that is  solved analytically and has a unique solution~\cite{Salenbauch2019}, while in the case of the EQMOM, an iterative and non-unique procedure~\cite{Nguyen2016, Pigou2018,Yuan2012}  is applied to invert the moments from low- and high-order moments of the entire NDF. 
\colorblue
It should be noted that although the S-EQMOM and the EQMOM share similarities, the moment inversion strategy in the two approaches is different, as described above, so the positions and weights determined in the  S-EQMOM do not necessarily coincide with the EQMOM values~\cite{Salenbauch2019}.
\colorblack
The S-EQMOM formulation greatly improves the stability of the inversion algorithm, allowing a computationally efficient and robust local reconstruction of the  soot  particle NDF~\cite{Salenbauch2019}. 

In order to capture both unimodal and bimodal particle distributions which arise from the interplay of particle nucleation and coagulation~\cite{Zhao2003}, a special treatment of these two source terms  was formulated and validated  in~\cite{Salenbauch2019}. 
The nucleation source term is included only in the first sub-NDF $n_{s_1}$, which tracks small, freshly formed particles. 
\colorblue
As discussed in~\cite{Bartos2017}, the first particle mode consists  of so-called nanoparticles, i.e. freshly nucleated  particles with a diameter of less than 10 nm. 
\colorblack
The other sub-NDFs are therefore not affected  by nucleation and can evolve to larger particle sizes. The coagulation source term takes into account collisions between particles not only from the same sub-NDF but also  from different sub-NDFs, proving a  coupling term between the sub-NDFs~\cite{Salenbauch2019}.
\colorblue
Further, condensation, chemical surface growth (HACA), and oxidation are processes active on all sub-NDFs.  
 The detailed formulation of the source terms is described in our previous work \cite{Salenbauch2019} and is not repeated here for brevity. 

The soot kinetics  employed  in this study  follows our previous works~\cite{Salenbauch2019, Salenbauch2015a} and it is here only shortly described. 
The particle nucleation is modeled by  a dimerization reaction of two PAH molecules~\cite{Balthasar2003}. A  lumped PAH species, $Y_{PAH}$,  is  defined as the sum of pyrene \ce{A4} and cyclopentapyrene \ce{A_4R_5}. Condensation  is described as the collision of PAH molecules with soot particles~\cite{Balthasar2003}, while the soot surface growth follows the H-abstraction-\ce{C2H2}-addition (HACA) mechanism from Frenklach and Wang~\cite{Frenklach1991,Frenklach1994}. The rates of the HACA mechanism are taken from Appel et al. \cite{Appel2000}. 
 The coagulation process  accounts for the transition between the continuum and the free molecular regime by applying a harmonic mean interpolation of the collision kernels of the two individual regimes~\cite{Kazakov1998}.  Finally, particle oxidation is assumed to occur through reactions with \ce{O2} and OH~\cite{Frenklach1994} with  the  rate parameters  taken from  Appel et al. (2000)~\cite{Appel2000}. 

\colorblack

\subsection{Combustion Model}
\label{sec:numFlamelet}
In the present study, the flamelet/progress variable (FPV) approach~\cite{Pierce2004,Ihme2005}  is employed, in which flamelet solutions are generated by solving the steady-state flamelet equations~\cite{Peters1986} with different stoichiometric scalar dissipation rates ranging from thermochemical equilibrium to quenching conditions.

The chemical reactions are described by the kinetic mechanism presented in~\cite{Blanquart2009,Narayanaswamy2010}. 
The thermochemical state is hence parametrized by the mixture fraction $Z$ and the scalar dissipation rate $\chi_{st}$, which is mapped to the progress variable $Y_C$. 
Here, $Y_C$ is  defined as $Y_C=Y_{CO_2}+Y_{CO}+Y_{H_2O}+Y_{H_2}$. 
Soot radiation effects are neglected in this study due to the very low soot volume fraction of the flame~\cite{Han2018}. 


\subsection{Coupled LES-FPV approach}
\label{sec:numLES}
The flow field is described by the Favre-filtered Navier-Stokes equations. The turbulence closure is achieved using the eddy viscosity hypothesis with the $\sigma$ model~\cite{Nicoud2011}. The model constant is determined using the dynamic procedure~\cite{Toda2011}. 
The suitability of the LES-FPV numerical framework for turbulent jet flames was previously shown in \cite{Hunger2017, Popp2015, Gierth2018,Wen2021d}. 
 The Favre-filtered transport equations of the mixture fraction and progress variable are given by
\begin{equation}
\label{eqn:Z}
	\frac{\partial}{\partial t}\,(\bar{\rho}\,\widetilde{Z}) 
  + \dpartial{}{\bm{x}}(\bar{\rho}\,\widetilde{\boldsymbol{u}}\,\widetilde{Z})
  = \dpartial{}{\bm{x}} \left[\bar{\rho}\left(D_{Z}+\frac{\nu_{\text{t}}}{\text{Sc}_{\text{t}}}\right)\,\dpartial{\widetilde{Z}}{\bm{x}}\right],
\end{equation}
\begin{equation}
\label{eqn:PV}
	\frac{\partial}{\partial t}\,(\bar{\rho}\,\widetilde{Y}_C) 
  + \dpartial{}{\bm{x}}(\bar{\rho}\,\widetilde{\boldsymbol{u}}\,\widetilde{Y}_C)
  = \dpartial{}{\bm{x}}\left[\bar{\rho}\left(D_{Y_C}+\frac{\nu_{\text{t}}}{\text{Sc}_{\text{t}}}\right)\,\dpartial{\widetilde{Y}_C}{\bm{x}}\right]
  + \overline{\dot{\omega}}_{Y_C}.
\end{equation}
Here, $D_{Z}$ and $D_{Y_C}$ denote the diffusion coefficient of the mixture fraction and progress variable, respectively, which are evaluated under the assumption of a unity Lewis number as  $D_{Z}= D_{Y_C}=D=\alpha$, with the thermal diffusivity $\alpha$ retrieved from the flamelet table. Furthermore, $\nu_t$ represents the turbulent  viscosity, $\text{Sc}_\text{t}$ is the turbulent Schmidt number set equal to 0.7~\cite{Jones2010} and $\overline{\dot{\omega}}_{Y_C}$ is the filtered source term of the progress variable, which is also obtained from the flamelet table.  The variance of the mixture fraction $\widetilde{Z''^2}$ is calculated using an algebraic equation following ~\cite{Pierce2004}.

The thermochemical state is retrieved using a normalized progress variable $\widetilde{C}$~\cite{Domingo2008}, defined as  
\begin{equation}
 \widetilde{C}=\frac{\widetilde{Y}_C - \widetilde{Y}_{C\text{,min}}(\widetilde{Z})}{\widetilde{Y}_{C\text{,max}}(\widetilde{Z}) - \widetilde{Y}_{C\text{,min}}(\widetilde{Z})}.
\end{equation}
Here, the minimum and maximum progress variable values are  functions of the mixture fraction and are determined from the tabulated flamelet solutions.  To take into account non-resolved fluctuations, a  $\beta$-shaped filtered density function (FDF) is assumed for the mixture fraction, whereas a $\delta$-FDF is applied for the progress variable. 
Hence, the Favre-filtered thermochemical state $\widetilde{\phi}$ is parametrized as $\widetilde{\phi}(\widetilde{Z},\widetilde{Z^{\prime\prime 2}},\widetilde{C})$. 

To model the slow PAH chemistry and the mass transfer  from the gas to the solid phase due to  soot  particle formation, a  transport equation for the Favre-filtered PAH mass fraction is solved following \cite{Mueller2012a},
\begin{equation}
\dpartial{\bar{\rho} \widetilde{Y}_\mathrm{PAH}}{t}+
\dpartial{\bar{\rho} \tilde{\bm{u}} \widetilde{Y}_\mathrm{PAH}}{\bm{x}}=
 \dpartial{}{\bm{x}}\left[\bar{\rho}\left(D+\frac{\nu_{\text{t}}}{\text{Sc}_{\text{t}}}\right)\,\dpartial{\widetilde{Y}_\mathrm{PAH}}{\bm{x}}\right]
 + \overline{\dot{\omega}}_\mathrm{PAH}.
\end{equation}
 %
Here   $\widetilde{Y}_\mathrm{PAH}$ is the sum of the PAH soot precursors considered (pyrene $A_4$ and cyclopentapyrene $A_4R_5$). 
Following~\cite{Mueller2012a,Ihme2008} the filtered source term  $\overline{\dot{\omega}}_\mathrm{PAH}$ is decomposed into three components: a chemical production term  $\dot{\omega}_\mathrm{PAH}^+$ that is independent of the PAH species concentration, a chemical consumption term    $\dot{\omega}_\mathrm{PAH}^-$ that is a linear function of the species concentration, and a consumption term representing the mass  transfer rate from the gas phase to soot $\dot{\omega}_\mathrm{s}$, which  is quadratic with the species concentration, 
\begin{equation}
\dot{\omega}_\mathrm{PAH} = \dot{\omega}_\mathrm{PAH}^+
+ \dot{\omega}_\mathrm{PAH}^-
+ \dot{\omega}_\mathrm{s}.
\end{equation}

\noindent The filtered source term is then  decomposed as 
\begin{equation}
\overline{\dot{\omega}}_\mathrm{PAH} = 
\overline{\dot{\omega}}_\mathrm{PAH}^{+ T}
+ \overline{\dot{\omega}}_\mathrm{PAH}^{- T} \left( \dfrac{\widetilde{Y}_\mathrm{PAH}}{\widetilde{Y}_\mathrm{PAH}^T} \right)
+ \overline{\dot{\omega}}_\mathrm{s}^T \left( \dfrac{\widetilde{Y}_\mathrm{PAH}}{\widetilde{Y}_\mathrm{PAH}^T} \right)^2,
\end{equation}
where the superscript $T$ indicates the value obtained from the flamelet table. 
It is important to note that the rate $\dot{\omega}_{s}^T$ can also be determined in the laminar flamelet calculation since it is only dependent on the gas-phase soot precursor concentration, therefore it is obtained here along with the thermochemical state. 

\section{Application to a turbulent flame}
\subsection{Experimental configuration}
The experimental setup under investigation is the Delft-Adelaide Flame III, a benchmark case of  the International Sooting Flame workshop~\cite{ISF}. The burner was first described in \cite{Peeters1994} and consists of a central main fuel jet with a diameter of 6 mm. 
This is surrounded by a rim containing 12 pilot flames stabilizing the main flame, a primary air annular coflow  with  inner and outer diameters equal to   15 mm and  45 mm, respectively, and a secondary  coflow. The fuel jet has a bulk velocity of  $21.94$ m/s and the coflow streams have a velocity  of $4.4$ m/s and $0.3$ m/s,  respectively. 
The fuel used in the experiments was commercial natural gas. Its composition varied  slightly between the experiments performed on this burner in different laboratories. However, the adiabatic flame temperature variations were small with respect to the variation in fuel composition~\cite{Nooren2000}. 
In this study, the  natural gas composition used for the soot particle measurements~\cite{Qamar2009} was chosen.
Laser-Doppler anemometry (LDA) data are available for the flow field~\cite{Stroomer1995}, while Raman-Rayleigh-LIF measurements provide the temperature, major species mass fractions, and mixture fraction~\cite{Nooren2000}. Soot measurements were performed by~\cite{Qamar2009} using planar laser-induced incandescence (LII).

\subsection{Numerical setup}
The numerical simulations are performed using the software OpenFOAM (v2006)~\cite{OpenFOAM} to solve the filtered  Navier-Stokes,  reactive scalar, and moment transport equations. In-house libraries are used for the combustion and turbulence-combustion closures~\cite{ulf,Weise2013} and for the S-EQMOM closure~\cite{Salenbauch2019}. 
In the S-EQMOM,  the NDF is described using $N_s=2$ sub-NDFs, which is a suitable number to approximate typical bimodal soot particle distributions, as shown by Salenbauch et al.~\cite{Salenbauch2019}. 
This yields  six additional transport equations for solving the first three moments of each sub-NDF. 

The computational domain is a cylinder  that spans 150 $D$ in the axial direction and 33.3 $D$ in the radial direction, with $D$ being the main fuel jet diameter. The domain is discretized by a block-structured  mesh with approximately 12.8 million hexahedral cells, with 1000, 184 and  68 cells in the axial, radial and circumferential directions, respectively. 
In the axial and circumferential directions, a uniform mesh is employed; 
in the radial direction, the mesh is refined close to the main jet, pilot and inner coflow and stretched towards the  lateral direction. 
For the generation of the turbulent inflow boundary conditions at the main jet and primary coflow, two LESs of the upstream geometry were performed separately.  
A convergent  pipe was calculated for the main jet and a divergent  annulus pipe for the primary coflow, according to the burner geometry~\cite{Nooren2000}. A bulk velocity was applied at the secondary coflow. 
Following Mueller and Pitsch~\cite{Mueller2012a} and Ayache and Mastorakos~\cite{Ayache2012}, the pilot nozzles are  modeled  as an annulus ring  with  a bulk velocity boundary condition, which preserves the experimental mass flow rate. 

An implicit second-order temporal discretization scheme is employed  with a maximum Courant number of 0.03 and a second-order central difference scheme is used for the convective flux in the momentum equations. A TVD scheme using the Sweby limiter~\cite{Sweby1984}  is applied for the convective scalar fluxes.

The simulation results discussed in Section~\ref{sec:results} were obtained by collecting temporal statistics over 0.15~s and  by averaging in the circumferential direction.  The simulations were computed on  576 cores on six Intel  Xeon Platinum 9242 nodes with a total computational cost of about 0.5 million CPU hours. 

\section{Results}
\label{sec:results}
In this section, the simulation results of the gas phase are first compared to the experimental data from~\cite{Stroomer1995,Nooren2000}. Velocity and reacting scalar data  are only available in the upstream region, where no soot is formed.  
The soot-related quantities  compared with LII data from~\cite{Qamar2009} are then presented and discussed, providing a validation of the S-EQMOM approach in LES of turbulent flames. The PSD, which is not available from the experiments,  is hence reconstructed and analyzed at different positions in the flame in order to investigate its spatial evolution. Finally, the soot dynamics and its relation with the gas-phase dynamics are examined to gain a further insight into the high soot  intermittency of this flame. 

\subsection{Gas-phase statistics}
\label{sec:gasphase}
\begin{figure}[h]
\centering
\includegraphics[scale=0.9]{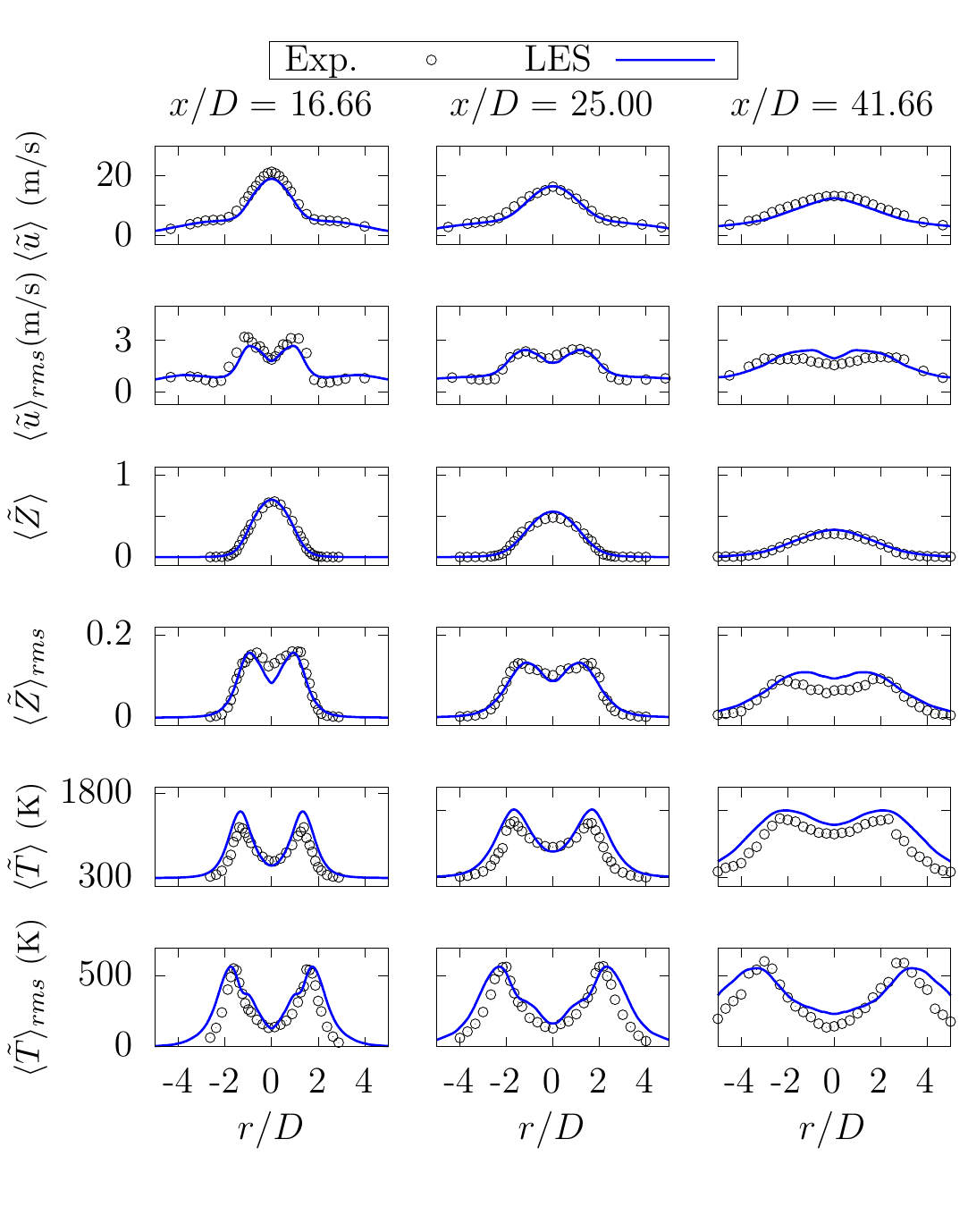}
\caption{Radial distributions of time-averaged axial velocity, mixture fraction and temperature together with the corresponding rms values at three axial locations above the burner. Solid line: LES; symbols: experimental data.}
\label{fig:gasPhaseUZT}
\end{figure}

The time-averaged radial distributions of the axial velocity, mixture fraction, and temperature and their root mean square (rms) are shown in Fig.~\ref{fig:gasPhaseUZT} at three axial locations: $x/D=$ 16.66, 25, and 41.66. 
The simulation results show very good agreement with the experimental data at all locations, both in terms of  mean and rms values, indicating that the flow field, scalar mixing and subsequent combustion are predicted accurately.
\colorblue
A slight overestimation is observed for  time-averaged temperature profiles. Using only steady flamelet solutions in the LES/FPV approach leads to an under-prediction of the local extinction in the first portion of the flame. 
More advanced LES/FPV models,  which include gas-phase radiation~\cite{Ihme2008c} or assume an extended presumed PDF for the reactive scalars~\cite{Ihme2008}, have shown improved prediction in flames with a comparable amount of local extinction, e.g., the Sandia Flame E. 
However, the over-prediction of the temperature is found to have a minor effect on the soot formation process downstream in the flame, as discussed below.     
\colorblack

\begin{figure}[h]
\centering
\includegraphics[scale=0.9]{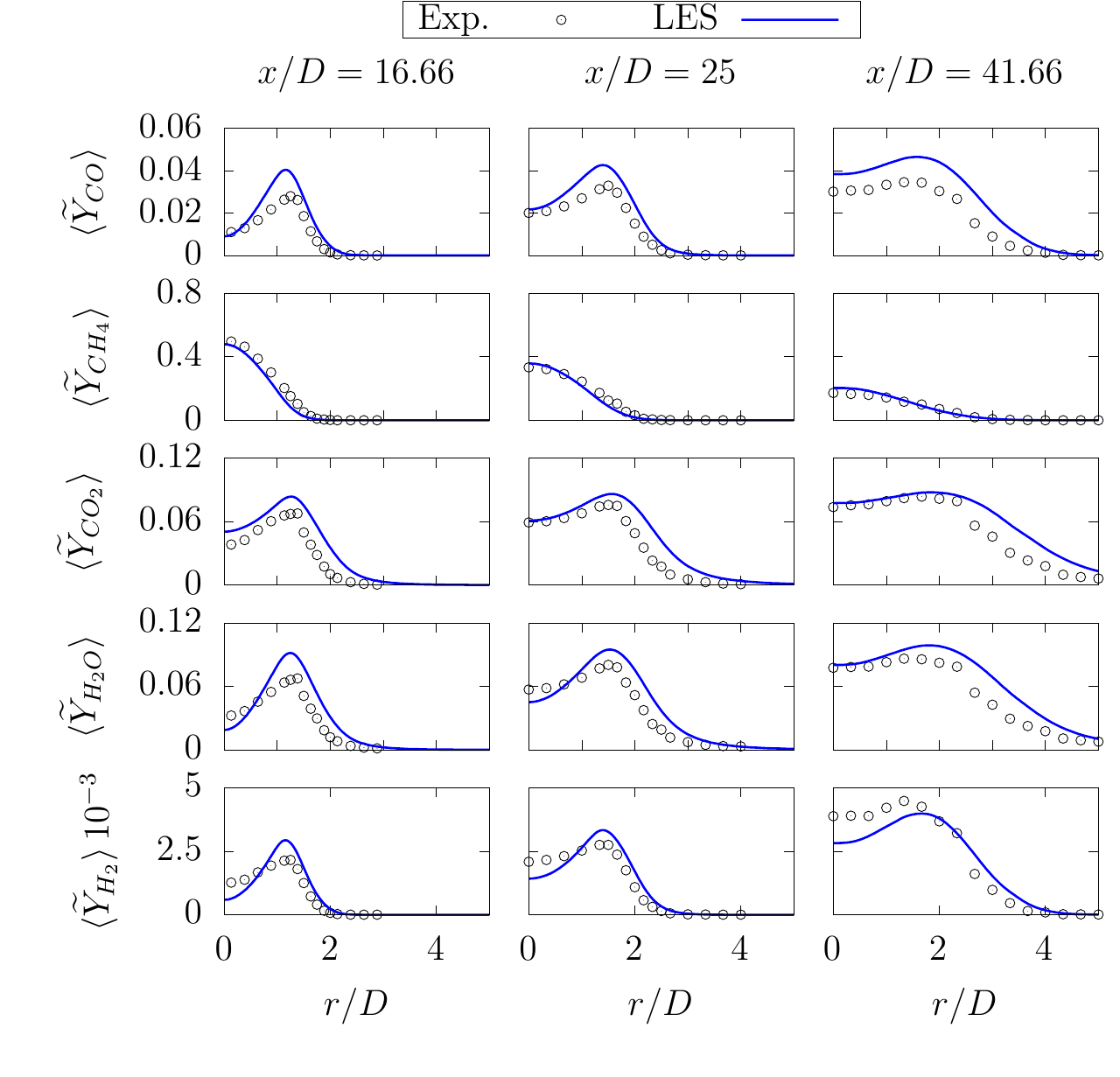}
\caption{Radial distributions of time-averaged \ce{CO}, \ce{CH4}, \ce{CO2}, \ce{H2O} and \ce{H2}  mass fractions at three axial locations above the burner. Solid line: LES; symbols: experimental data.}
\label{fig:gasPhaseSpecies}
\end{figure}

Figure~\ref{fig:gasPhaseSpecies} shows the time-averaged radial profiles of the \ce{CO}, \ce{CH4}, \ce{CO2}, \ce{H2O} and \ce{H2}   mass fraction. Similarly, the LES results indicate that the experimental species data are estimated well at all locations.
This further shows that the turbulence and turbulent combustion models applied are able to correctly capture the main features of the flame structure in the gas phase. 

\FloatBarrier
\subsection{Soot-phase statistics}
\label{sec:soot_results}
\begin{figure}[h]
\centering
\includegraphics[width=0.7\textwidth,trim=0.1cm 0cm 0.1cm 0cm,clip]{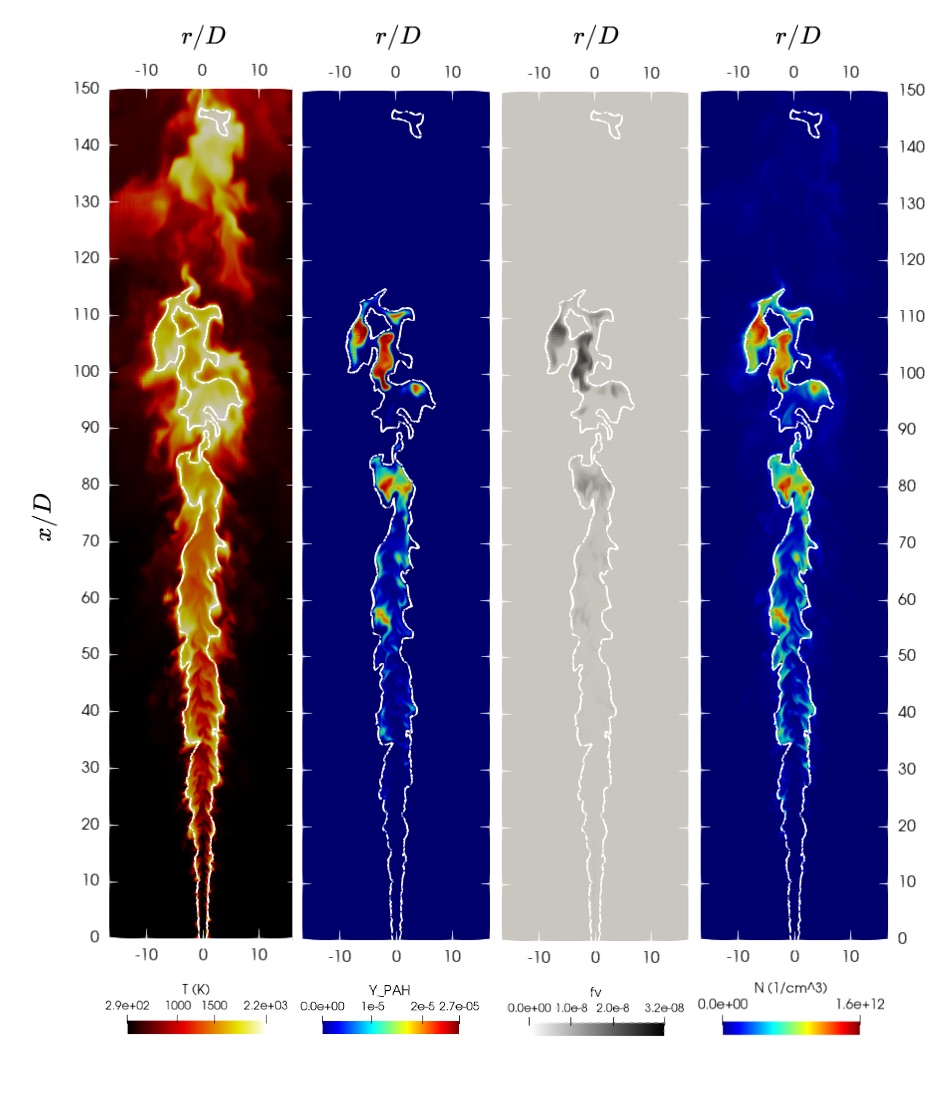}
\caption{Instantaneous contours of temperature $T$, \ce{PAH} mass fraction $Y_\mathrm{PAH}$, soot volume fraction $f_V$ and particle number density $N$. The white line represents the stoichiometric mixture fraction isoline  $Z=Z_{st}=$  0.073.}
\label{fig:snapshot}
\end{figure}

Instantaneous contours of the temperature, PAH mass  fraction, soot volume fraction, and particle number density  are shown in Fig.~\ref{fig:snapshot}. The superimposed  white line represents the stoichiometric mixture fraction isoline ($Z=Z_{st}=0.073$). 
It can be observed that the soot precursor mass fraction $Y_\mathrm{PAH}$ is produced on the fuel-rich side of the flame. 
A significant amount of soot is formed downstream  in the flame, for $x/D>$ 50.   Due to turbulent fluctuations,  fuel-rich pockets sporadically detach from the main flame zone,  connected to the fuel-rich core,  at around $x/D=$ 80–100, and are transported further downstream where they are  oxidized.  
This flame region is characterized by high soot intermittency, as observed in the experiments in~\cite{Qamar2009} and previous numerical studies~\cite{Mueller2012a, Han2018}. This phenomenon will  be discussed in more detail in Section~\ref{sec:intermittency}. 

\begin{figure}[h]
\centering
\includegraphics[width=0.7\textwidth,trim=0cm 0cm 0.1cm 0.5cm,clip]{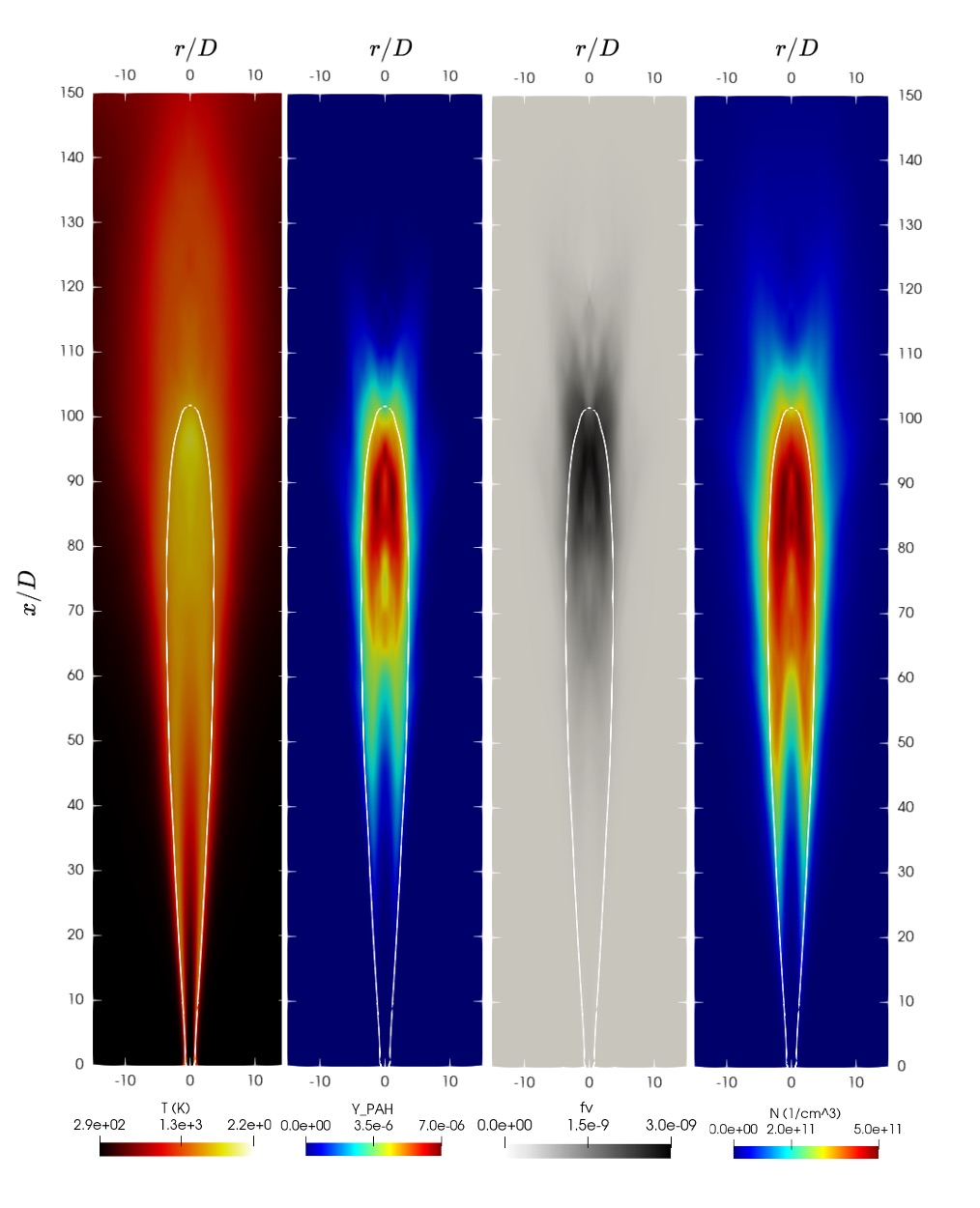}

\caption{Time-averaged contours of temperature $T$, \ce{PAH} mass fraction $Y_\mathrm{PAH}$, soot volume fraction $f_V$ and particle number density $N$. The white line represents the stoichiometric mixture fraction isoline  $Z=Z_{st}=$ 0.073.}
\label{fig:snapshotMean}
\end{figure}

Time-averaged contours of the same quantities are shown in Fig.~\ref{fig:snapshotMean}. 
The number density contour indicates the presence of a  significant number of particles, $\mathcal{O}(10^{11}$ 1/cm$^3)$, between $x/D=$ 30 and 120, while the soot volume fraction is of the order of particles per billion (ppb) between $x/D=$~50 and 125, which is also beyond the time-averaged stoichiometric mixture fraction isoline. 

\begin{figure}[h]
\centering
\includegraphics[width=0.7\textwidth,trim=0.3cm 0cm 0.1cm 0.1cm,clip]{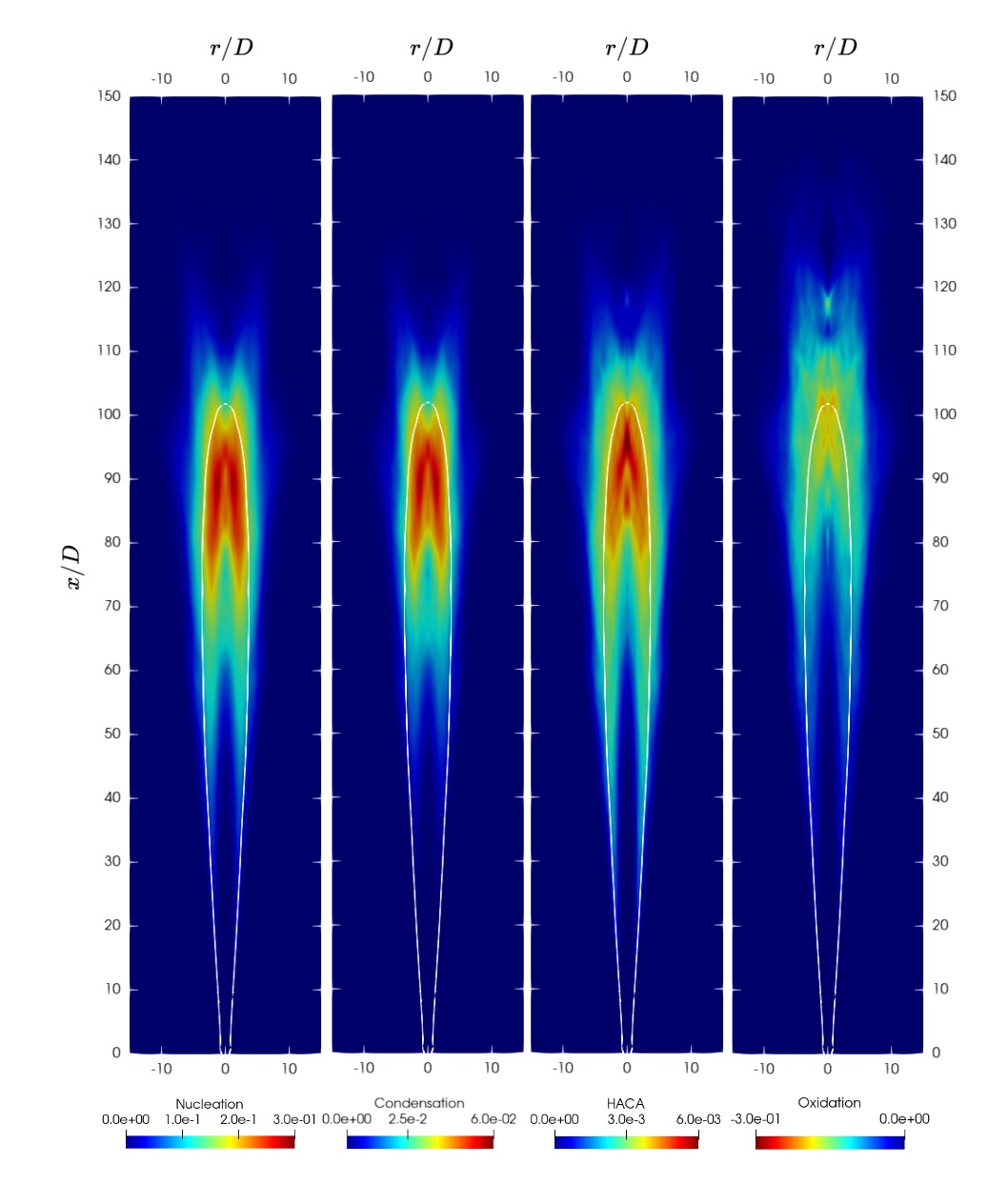}

\caption{Time-averaged contours of  volume fraction source terms (ppm/s). The white line represents the stoichiometric mixture fraction isoline $Z=Z_{st}=$ 0.073. Note that  different color scales are used.}
\label{fig:snapshotMeanSourceTerms}
\end{figure}

The time-averaged volume fraction source terms are plotted in Fig.~\ref{fig:snapshotMeanSourceTerms}. Note that only  nucleation and oxidation rates are plotted on the same color scale, while  condensation and HACA rates are plotted on different color scales due to their different orders of magnitude. 
It can be observed that nucleation and condensation  are the predominant processes on the fuel-rich side of the flame, although the maximum condensation rate is smaller than the nucleation rate by a factor of 5. The HACA process is instead one order of magnitude smaller than the condensation, as also numerically observed in~\cite{Mueller2012a}. It  takes place in regions with a locally rich mixture, reaching a maximum in the region   between  $x/D=$~80 and 100, close to the flame tip,  and also, with minor intensity,  around the stoichiometric isoline. 
The oxidation rate is of the same order of magnitude as the nucleation and takes place around the stoichiometric isoline  and downstream in the lean flame region, where most of  the soot particles are oxidized.   


\FloatBarrier 

\begin{figure}[h!]
\centering
\begin{subfigure}[h]{0.4\textwidth}
\includegraphics[scale=0.9]{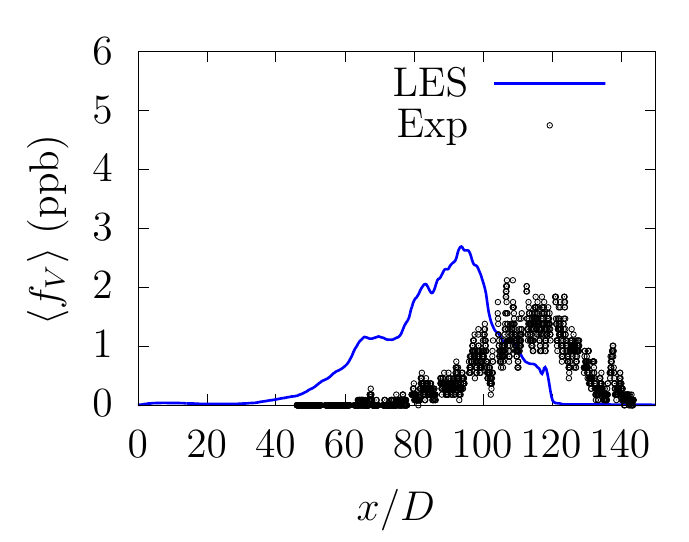}
\caption{}
\label{fig:fv_centerlineA}
\end{subfigure}
\begin{subfigure}[h]{0.4\textwidth}
\includegraphics[scale=0.9]{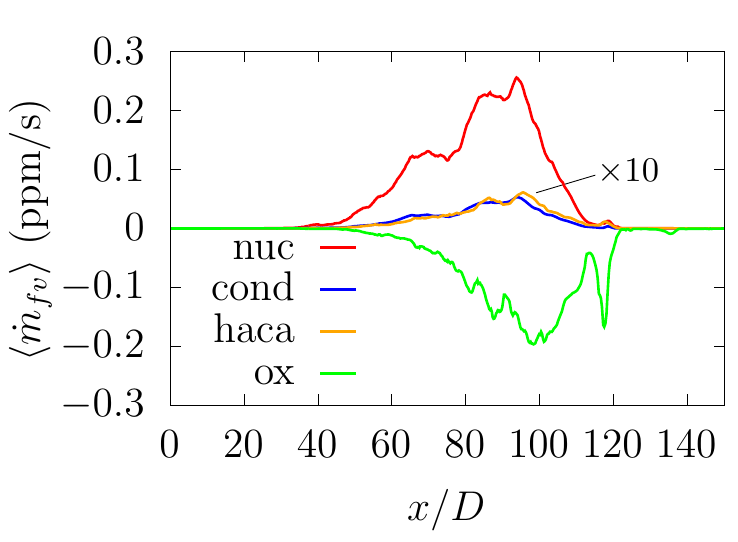}
\caption{}
\label{fig:fv_centerlineB}
\end{subfigure}
\caption{(a): Time-averaged soot volume fraction on the centerline. Solid line: LES; symbols: experimental data from Qamar et al.~\cite{Qamar2009}. (b): Time-averaged soot volume fraction source terms on the centerline. The HACA source term is multiplied by 10.}
\label{fig:fv_centerline}
\end{figure}

A quantitative comparison is shown in Fig.~\ref{fig:fv_centerlineA}, where the evolution of the time-averaged experimental and numerical soot volume fraction along the flame centerline is plotted. It can be observed that the maximum value is  slightly over-predicted and  is located at $x/D=$~97, upstream compared to  the maximum position from the  experimental data at $x/D\approx$~115.  
The complete oxidation of soot particles is also  predicted  upstream, at $x/D=$~125,  against $x/D=$ 140 in the experiments.  However, compared to previously  published results for the same flame using different modeling approaches~\cite{Donde2012a,Sewerin2018,Schiener2018},  the present results show a significant improvement in the prediction of the peak value and location of the experimental soot volume fraction. A similar level of agreement with the results reported in~\cite{Mueller2012a, Han2018}  is instead  observed. 
\colorblue
As pointed out  by Mueller and Pitsch~\cite{Mueller2012a}, the discrepancy in the  location of soot onset between simulation results and experiments may be due to the significant uncertainty  in  PAH chemistry, which becomes  particularly relevant in methane/air flames with low sooting tendency. 

Furthermore, minor dependency of the particle formation  is observed concerning the temperature history along the flame. 
Comparing the LES/FPV results presented here with those obtained from an LES sparse multi
mapping conditioning~\cite{Huo2022a} or from an LES transported PDF~\cite{Han2018}, which can capture significant amount of
local extinction~\cite{Ge2013,Raman2007a} (see also Sec. \ref{sec:gasphase}), an earlier onset of soot particle formation is similarly predicted.

    
\colorblack
Figure~\ref{fig:fv_centerlineB} shows the time-averaged  soot volume fraction source terms on the centerline. The curves clearly indicate that nucleation and condensation are the predominant processes contributing to the soot particle formation and growth in this flame. 
Further, the oxidation source term plays an important role in the reduction of the soot volume fraction, being of the same order of magnitude as the nucleation  process  between $x/D=$ 60 and 100. Downstream, for $x/D>$ 100, oxidation is the primary process occurring on  soot particles, which is consistent with the qualitative observation discussed above.  
 


\begin{figure}[h]
\centering
\begin{subfigure}[h]{\textwidth}
\centering
\includegraphics[scale=0.8]{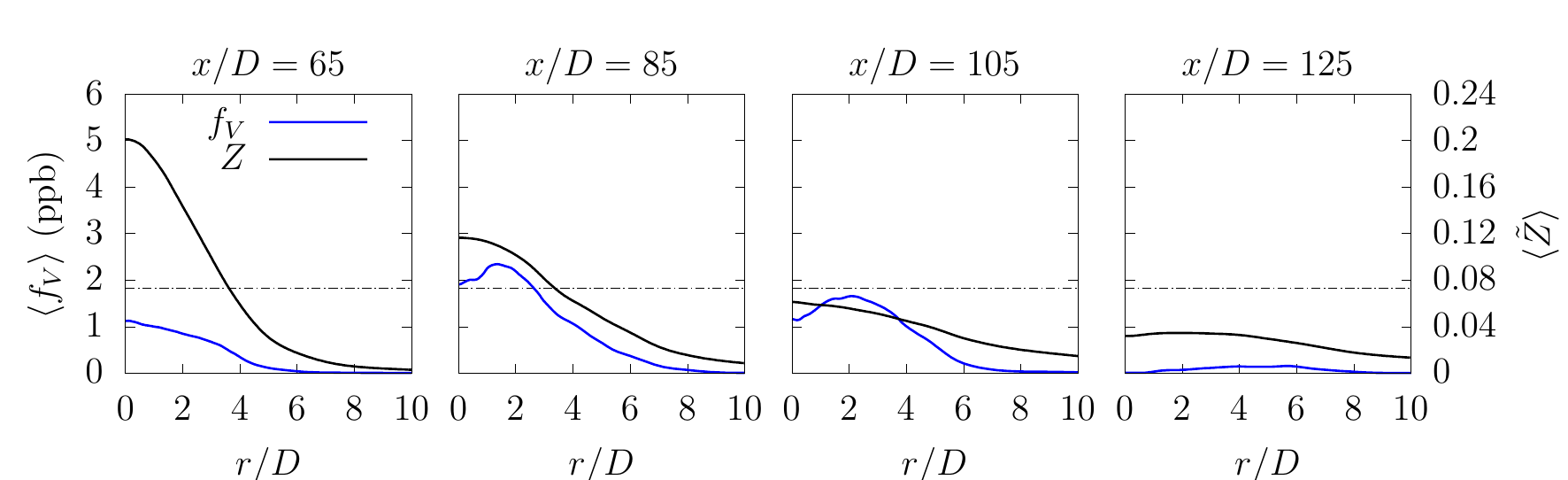}
\caption{}
\label{fig:sootRadialPlot} 
\end{subfigure}
\begin{subfigure}[[h]{\textwidth}
\centering
\includegraphics[scale=0.8,trim=0cm 0cm -0.8cm 0cm,clip]{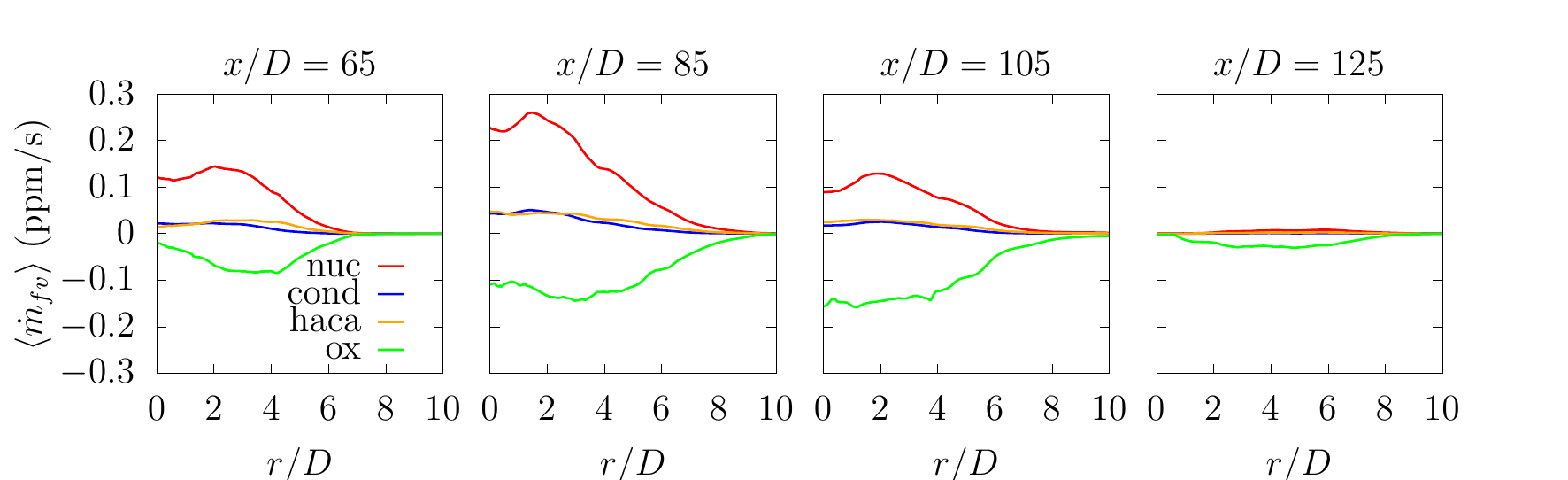}
\caption{}
\label{fig:sootRadialSourceTerm} 
\end{subfigure}
\begin{subfigure}[[h]{\textwidth}
\centering
\includegraphics[scale=0.8,trim=0cm 0cm -0.8cm 0cm,clip]{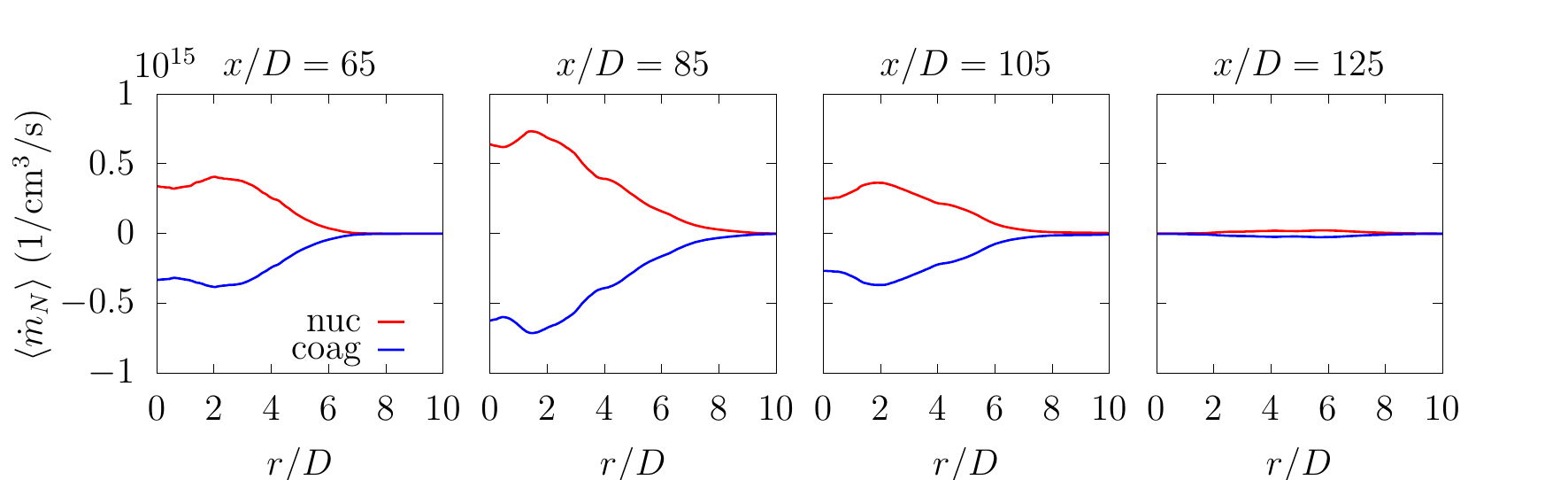}
\caption{}
\label{fig:sootRadialSourceTerm_m0} 
\end{subfigure}

\caption{(a): Radial profiles of the time-averaged soot volume fraction and mixture fraction at four axial locations. The horizontal dashed line represents the stoichiometric mixture fraction $Z=Z_{st}=$~0.073. \colorblue (b): Radial profiles of the time-averaged soot volume fraction source terms at four axial locations. The HACA source term is multiplied by 10. (c): Radial profiles of the time-averaged particle number source terms for nucleation and coagulation at four axial locations.}
\label{fig:radialSootComplete}
\end{figure}

\colorblue
Figure~\ref{fig:radialSootComplete}  shows the evolution of the time-averaged soot volume fraction and the mixture fraction along the radial coordinate (top), the  time-averaged soot volume fraction source terms (center), and the time-averaged number density source terms (bottom). 
\colorblack
 The profiles in Fig.~\ref{fig:sootRadialPlot} indicate that the soot volume fraction increases between $x/D=$ 65 and 85, while it is almost zero at $x/D=$ 125, where the particles are almost completely oxidized. 
At  $x/D=$~85, close to the position of the maximum soot volume fraction  on the centerline (see Fig. \ref{fig:fv_centerlineA}),  the profile reaches a maximum  at $r/D =$~2, where nucleation, condensation and HACA  are  concurrent processes,
\colorblue
 and the nucleation is close to its maximum value,  as shown in Fig.~\ref{fig:sootRadialSourceTerm}. 
The coagulation source term in Fig.~\ref{fig:sootRadialSourceTerm_m0} also reaches  at $x/D=$~85 its maximum  value. 
At $x/D=$ 105, the oxidation rate  is close to its maximum value and  is the dominant process at all radial locations. 
\colorblack
 
Furthermore, in all the  plots in Fig.~\ref{fig:sootRadialPlot}, the stoichiometric mixture fraction is indicated by the horizontal dashed line. 
It is shown that   at all axial locations, time-averaged soot volume fraction is present at radial locations where the time-averaged mixture fraction is below the stoichiometric value. 

In summary, the  results presented here are in good agreement with the experimental data and  comparable with the state-of-art soot simulations for this turbulent flame~\cite{Mueller2012a,Han2018} for the time-averaged values.  
The good prediction of the experimental soot quantities is  a prerequisite of the soot model to provide further insights into the undergoing physico-chemical processes.
Therefore,  in the following, the detailed simulation results are used to better understand the formation and evolution of the soot particles along the flame and their relationship with the gas-phase mixture.

\begin{figure}[h]
\centering
\includegraphics[scale=0.8]{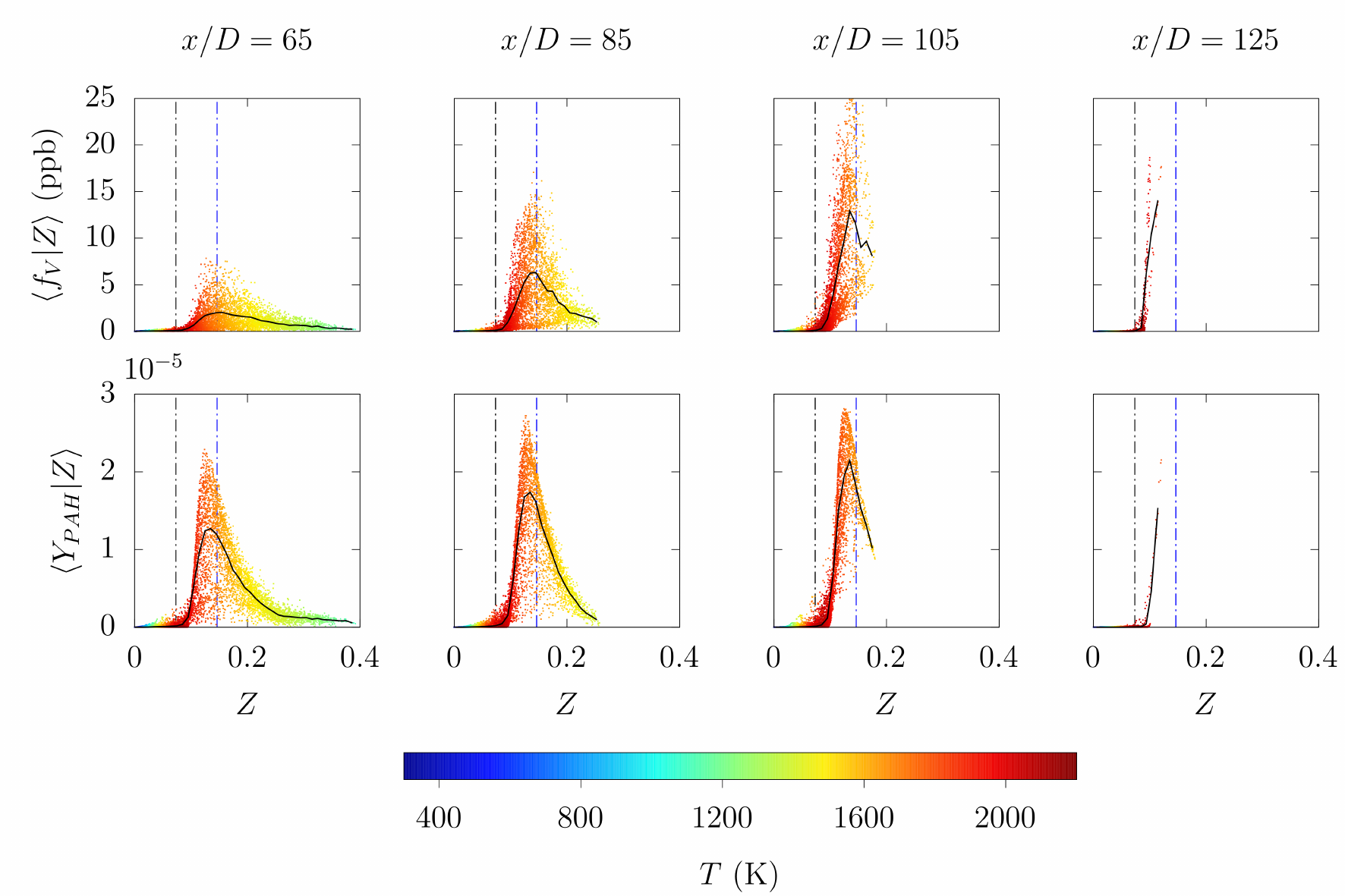}
\caption{Time-averaged mixture fraction conditional mean and scatter plots of the soot volume fraction (top) and PAH mass fraction (bottom) colored according to the temperature at  four axial locations. The black and blue vertical dashed lines indicate the locations of the stoichiometric mixture fraction $Z=Z_{st}=0.073$ and $Z=2Z_{st}=0.146$.}
\label{fig:scatterplot}
\end{figure}

The soot quantities are now investigated in mixture fraction space at different positions along the flame.
Figure~\ref{fig:scatterplot} shows  the conditional  scatter plots  of the soot volume fraction and \ce{PAH} mass fraction  versus the mixture fraction.  
The samples  represent instantaneous values collected over several instants in time, colored according to their temperature. 
The conditional means are also shown by the black solid line. Furthermore, the dashed black and blue lines indicate the stoichiometric mixture fraction $Z=Z_{st}=0.073$ and  $Z=2Z_{st}=0.146$, respectively. 
The soot volume fraction and PAH mass fraction are seen to exhibit a similar dependency on the mixture fraction. 
High values of soot volume fraction and PAH mass fraction are obtained at  mixture fraction values between $Z_{st}$ and $2Z_{st}$ and high temperatures. 
Conditional  soot volume fraction and $Y_\mathrm{PAH}$ profiles similarly reach a maximum  in this mixture fraction range 
at all the axial locations examined. 
Most of the samples are located on the rich mixture fraction side; the PAH mass fraction  rapidly  decreases on the lean mixture fraction side, similarly to the soot volume fraction. From $x/D=$~65 to  $x/D=$~105, samples with a higher soot volume fraction are detected,   but they span  a smaller range on the rich mixture fraction side, similarly to the PAH mass fraction. 
At $x/D=$ 125, downstream of  the location of maximum oxidation, only few samples are detected with high  volume fraction values and temperatures at mixture fractions close to stoichiometry.  

Furthermore, although  the time-averaged contours and profiles in Figs.~\ref{fig:snapshotMean} and~\ref{fig:sootRadialPlot} indicated the presence of soot in regions with a lean time-averaged mixture fraction, the conditional scatter plots show that instantaneous  samples with a significant soot volume fraction are found only for rich mixture fraction.  

\begin{figure}[h!]
\centering
\includegraphics[width=0.95\textwidth]{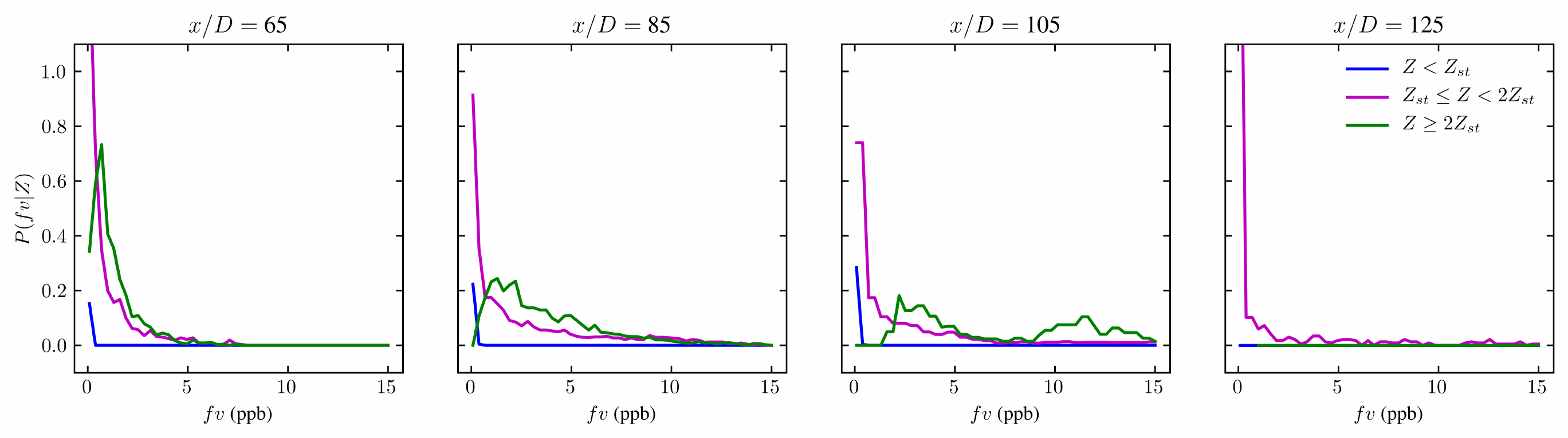}
\caption{Time-averaged  soot volume fraction PDF conditioned on the mixture fraction at different axial locations along the flame.}
\label{fig:condPDF_Z}
\end{figure}

To further analyze the correlation between the mixture fraction and soot volume fraction, 
the conditional PDFs of the soot volume fraction on the mixture fraction, $P(f_V\vert Z)$, are investigated.  
$P(f_V\vert Z)$, calculated based on the same samples used in Fig.~\ref{fig:scatterplot},   is  shown in Fig.~\ref{fig:condPDF_Z}. Three mixture fraction intervals are examined:  lean, with $Z<Z_{st}$, slightly rich, with $Z_{st}<Z<2Z_{st}$ and highly rich, with $Z>2Z_{st}$. 
The plots indicate that on the lean mixture fraction side (blue line), the conditional  soot volume fraction PDF has a peak at $f_V=$ 0 ppb and  is only not zero for  values of $f_V<$ 1~ppb along the whole length of the flame. 
This confirms that there is only a small  number of sooting samples at  lean condition and these disappear rapidly due to oxidation. 
For slightly rich mixtures, the conditional PDF has a peak at $f_V=$ 0 ppb followed by a rapid decay towards zero. 
For highly rich mixture fraction values, the conditional PDF has a peak for $f_V>0$ ppb and  becomes broader close to  $x/D=85$. 
Moreover, at $x/D=$ 105 the conditional PDF shows a bimodal shape, with 3 ppb and 12 ppb being the peak positions, respectively. 
This suggests the presence of samples containing lower/higher levels of $f_V$ that may correspond to   smaller/larger particles. Indeed, it is only at very rich conditions and high temperature that small particles can undergo significant surface growth due to the concurrent condensation and HACA processes. 
 At $x/D=125$, only the volume-fraction-conditional PDF at the slightly rich mixture fraction ($Z_{st}<Z<2Z_{st}$) is other than zero, and  samples with a very small soot volume fraction, $f_V<2$ ppb,  are detected at this position, while the volume fraction has been completely oxidized on the lean mixture fraction side.  

In conclusion, the results presented above illustrate that soot particle formation and growth occur only on the rich mixture fraction side, while soot is rapidly oxidized near the stoichiometric mixture fraction isoline. 
Therefore, the soot volume fraction present  in the time-averaged field beyond the stoichiometric mixture fraction isoline may be due  to the high soot intermittency observed experimentally in this portion of the flame. 
This will be analyzed in detail in Section~\ref{sec:intermittency}, while the next section discusses the reconstructed PSD obtained using the S-EQMOM. 


\FloatBarrier




   


\subsection{Mean Particle Size Distribution}
As stated above, the S-EQMOM method  offers the  significant advantage of providing information for the temporal and  spatial  evolution of the reconstructed particle size distribution.
In this section, the time-averaged  reconstructed soot PSD is investigated at different locations in the flame. 

\begin{figure}[h]
\centering
\includegraphics[scale=1]{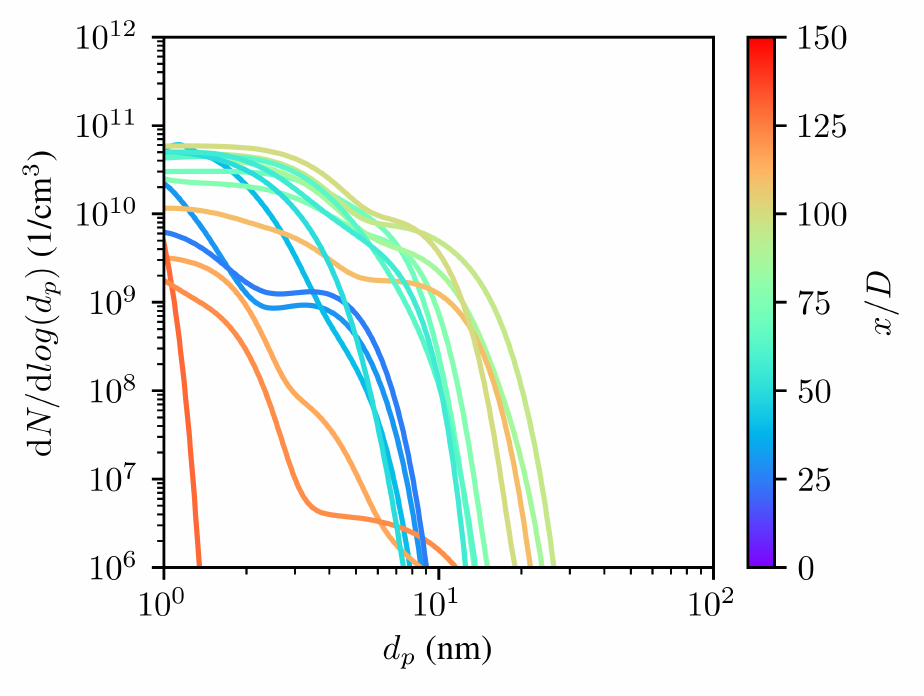} 
\caption{Time-averaged PSD of soot particles at different positions along the centerline.}
\label{fig:PSD_centerline}
\end{figure}

The  time-averaged  PSD along the flame centerline is plotted in Fig.~\ref{fig:PSD_centerline}. Upstream of $x/D=$~25, the PSD is not detected, which is consistent with the mean average soot volume fraction discussed in Section~\ref{sec:soot_results} (see Fig.~\ref{fig:fv_centerlineA}).
Near $x/D =$ 40 (light blue curves), the PSD appears unimodal and indicates the presence of particles smaller than 10~nm.  
Between $x/D=$ 40 and $x/D=$ 70, the number density of particles  with a  diameter larger than 10 nm increases, while the number density of small particles remains almost constant, revealing strong  particle nucleation and surface growth in this portion of the flame. 
Between  $x/D=$ 75  and $x/D=$ 90,  the shape of the PSD  changes from a unimodal  to  a bimodal distribution, with a further increase in the number density of  large particles. 
For $x/D>$ 100, the number density 
 drops dramatically over all particle sizes due to  the oxidation process, which is dominant in the lean mixture.  
 
\begin{figure}[h]
\centering
\begin{subfigure}[b]{0.495\textwidth}
\centering
\includegraphics[width=\textwidth]{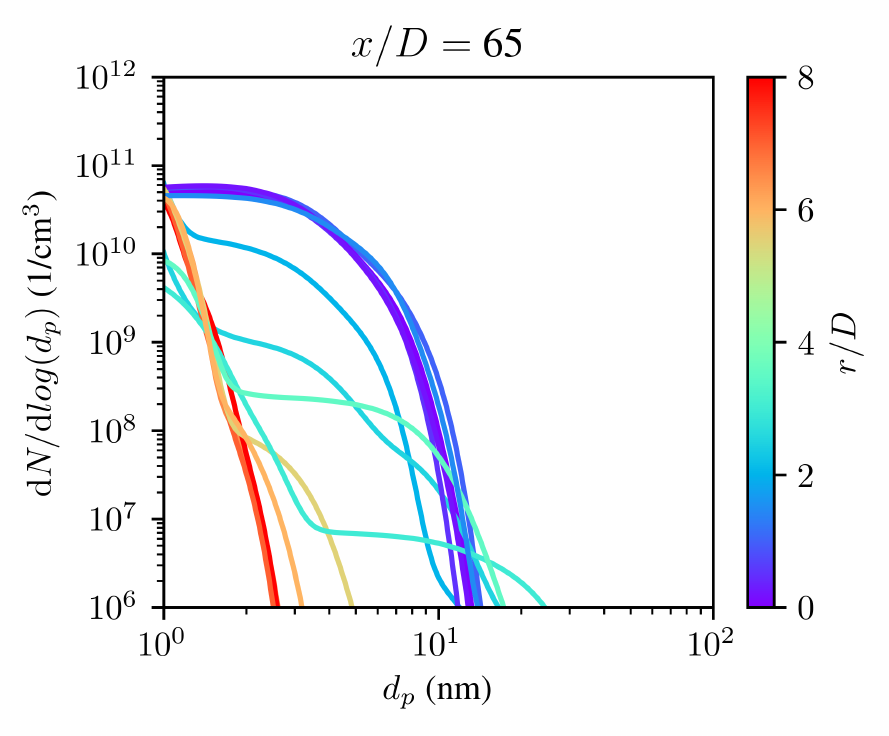}
\end{subfigure}
\hfill
\begin{subfigure}[b]{0.495\textwidth}
\centering
\includegraphics[width=\textwidth]{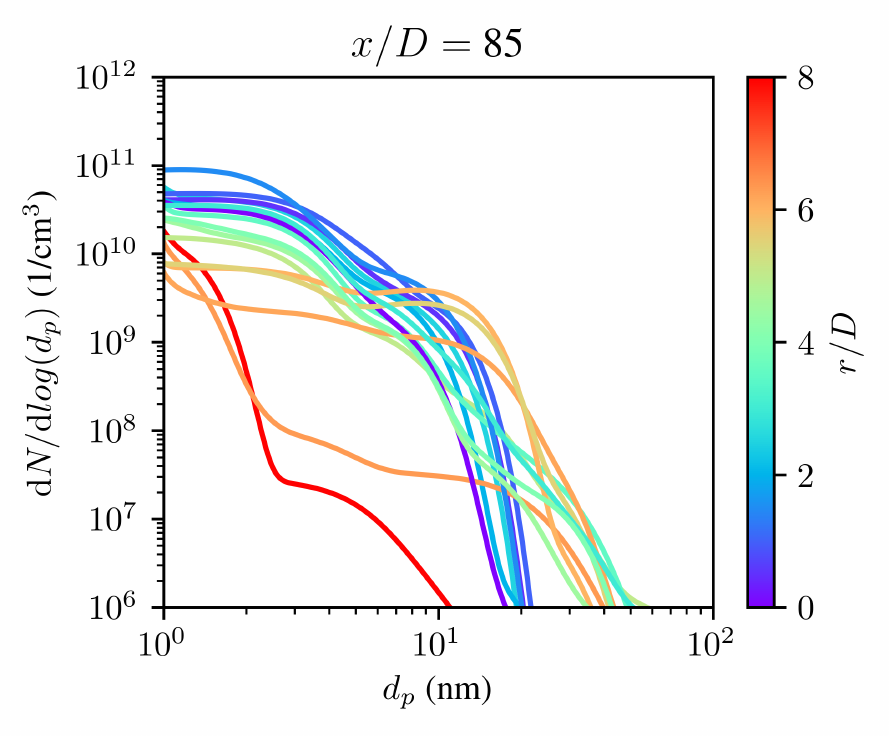}
\end{subfigure}
\begin{subfigure}[b]{0.495\textwidth}
\centering
\includegraphics[width=\textwidth]{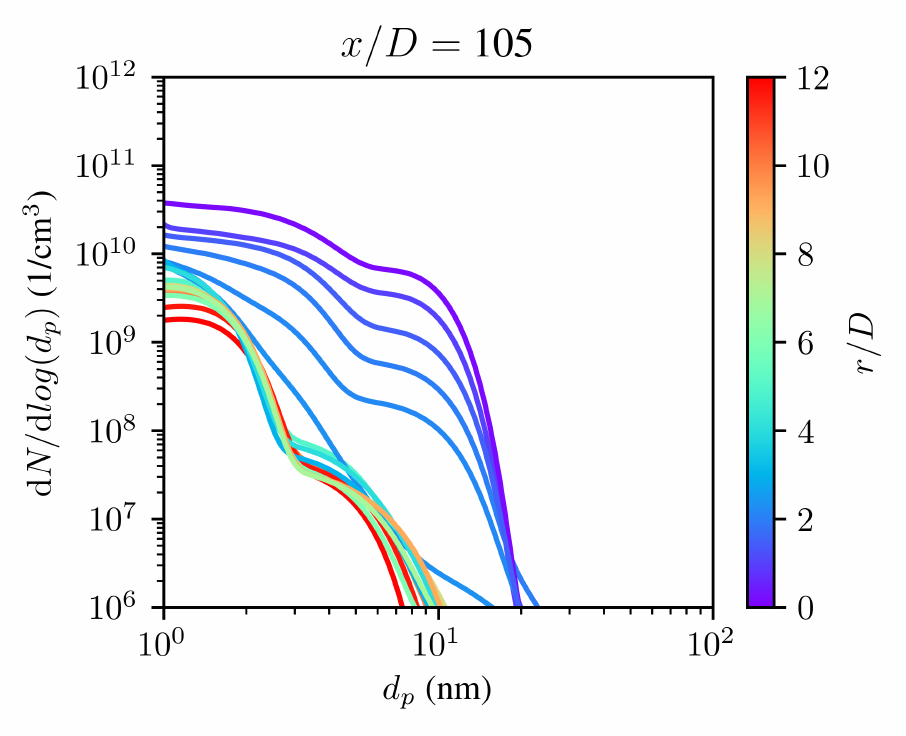}
\end{subfigure}
\caption{Time-averaged PSD of soot particles along the radius at  axial  position $x/D=$ 65 (top left), $x/D=$ 85 (top right) and $x/D=$ 105 (bottom).} 
\label{fig:PSD_radius}
\end{figure}

The time-averaged PSD evolution along the radial coordinate  is shown at three axial positions $x/D=$~65,  85 and 105  in Fig.~\ref{fig:PSD_radius}. The results indicate  a dominant unimodal distribution at  $x/D=$~65. 
At $x/D=$ 85, the PSD presents a bimodal shape between $r/D=$ 2 and 6, while the soot particles are oxidized in the case of larger radii. 
\colorblue
At this location, near the peak of the soot volume fraction, all soot processes are active and the strong coagulation observed in Fig.~\ref{fig:sootRadialSourceTerm_m0} contributes to the  transition of the PSD from unimodal  to bimodal.  
\colorblack
Downstream, at $x/D=$~105,  the PSD presents a  bimodal shape close to the centerline.
Its evolution outwards in the radial direction   illustrates a rapid decay of  the particle density at very small diameters and a shift in the PSD towards smaller particle diameters. 
Furthermore, in terms of both axial and radial PSD evolution,  the particle diameter remains significantly below 100 nm. 

Of the previous studies investigating this flame, only two analyzed the PSD~\cite{Sewerin2018,Huo2022a}. 
In~\cite{Sewerin2018},  a joint scalar-discrete number density PDF is solved using Eulerian stochastic fields, and  unimodal distributions are predicted along the centerline.
 In contrast, in \cite{Huo2022a}, where a sparse multiple mapping condition method   with a sectional method is used, a shift to a bimodal distribution is predicted near $x/D =$ 50,  slightly upstream compared to the results presented in this study.  The soot particle size was predicted to be smaller than 100 nm, similarly to the S-EQMOM results illustrated in this study.  

Since no experimental data are available for the PSD, the general  agreement with the work by Huo et al.~\cite{Huo2022a} is a promising result for  future applications of the S-EQMOM.
Further, it is noted that the computational cost of the S-EQMOM is lower than that in~\cite{Sewerin2018,Huo2022a}.




\FloatBarrier
\subsection{Dynamics of soot formation and oxidation}
\label{sec:intermittency}
As mentioned above, this flame is characterized by high soot intermittency. 
In the experimental work by Qamar et al.~\cite{Qamar2009}, the intermittency was defined  as the probability of not finding any soot (or below the measurement threshold of  0.1 ppb) at a given location and at a given time.  The  results indicated high soot intermittency for  $x/D>$ 80, with a peak at $x/D=$~110 (see Fig.~14 in ~\cite{Qamar2009}). 

\begin{figure}[h]
\centering
\includegraphics[scale=0.9]{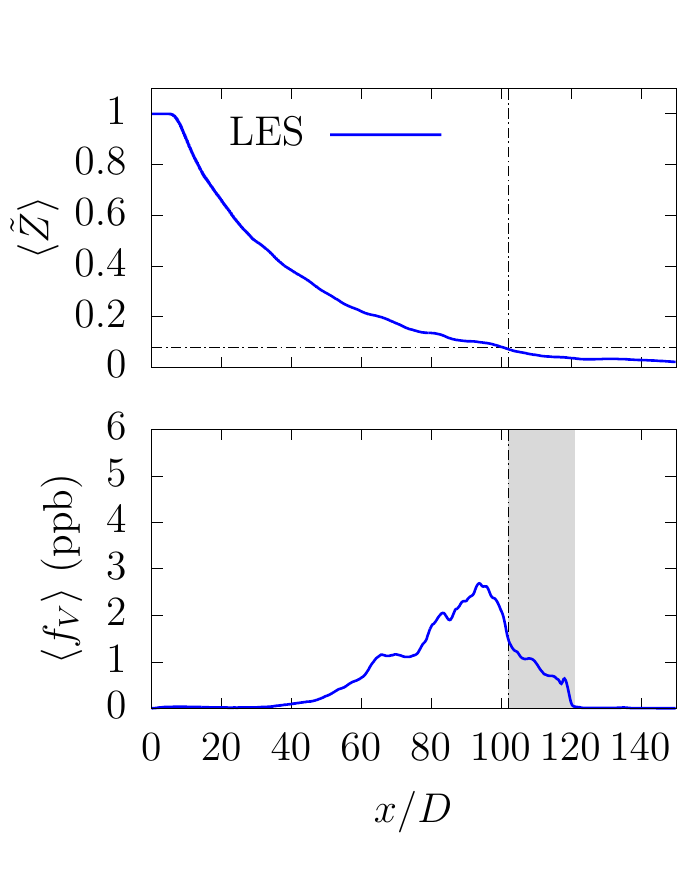}
\caption{Time-averaged mixture fraction and soot volume fraction profile  on the centerline. The horizontal  and vertical dashed lines indicate the stoichiometric mixture fraction value $Z=Z_{st}=$~0.073 and its axial location, respectively. The gray surface highlights the lean-mixture fraction side where the soot volume fraction is above zero.}
\label{fig:soot_Z_centerline}
\end{figure}

In order to further understand this phenomenon and its correlation with the underlying thermo-chemical state of the mixture,
 the time-averaged   mixture fraction and  soot volume fraction are first plotted
 along the centerline  in Fig.~\ref{fig:soot_Z_centerline}. 
 The soot volume fraction profile from Fig.~\ref{fig:fv_centerlineA} is repeated here for convenience.  The profiles show that soot particles do exist downstream of the axial position $x/D=$~102, where the time-averaged  mixture fraction reaches its stoichiometric value (indicated by the dashed line), but they are rapidly oxidized between $x/D =$ 102 and 125. This area with a lean time-averaged mixture fraction  is highlighted in gray in the bottom plot of  Fig.~\ref{fig:soot_Z_centerline}.  
The analysis in the following is therefore devoted to gaining further  insights into the soot quantities and their instantaneous and statistical behavior related to the mixture fraction field.    

\begin{figure}[h]
\centering
\includegraphics[width=0.99\textwidth,trim=0.2cm 0.2cm 0.1cm 0cm,clip]{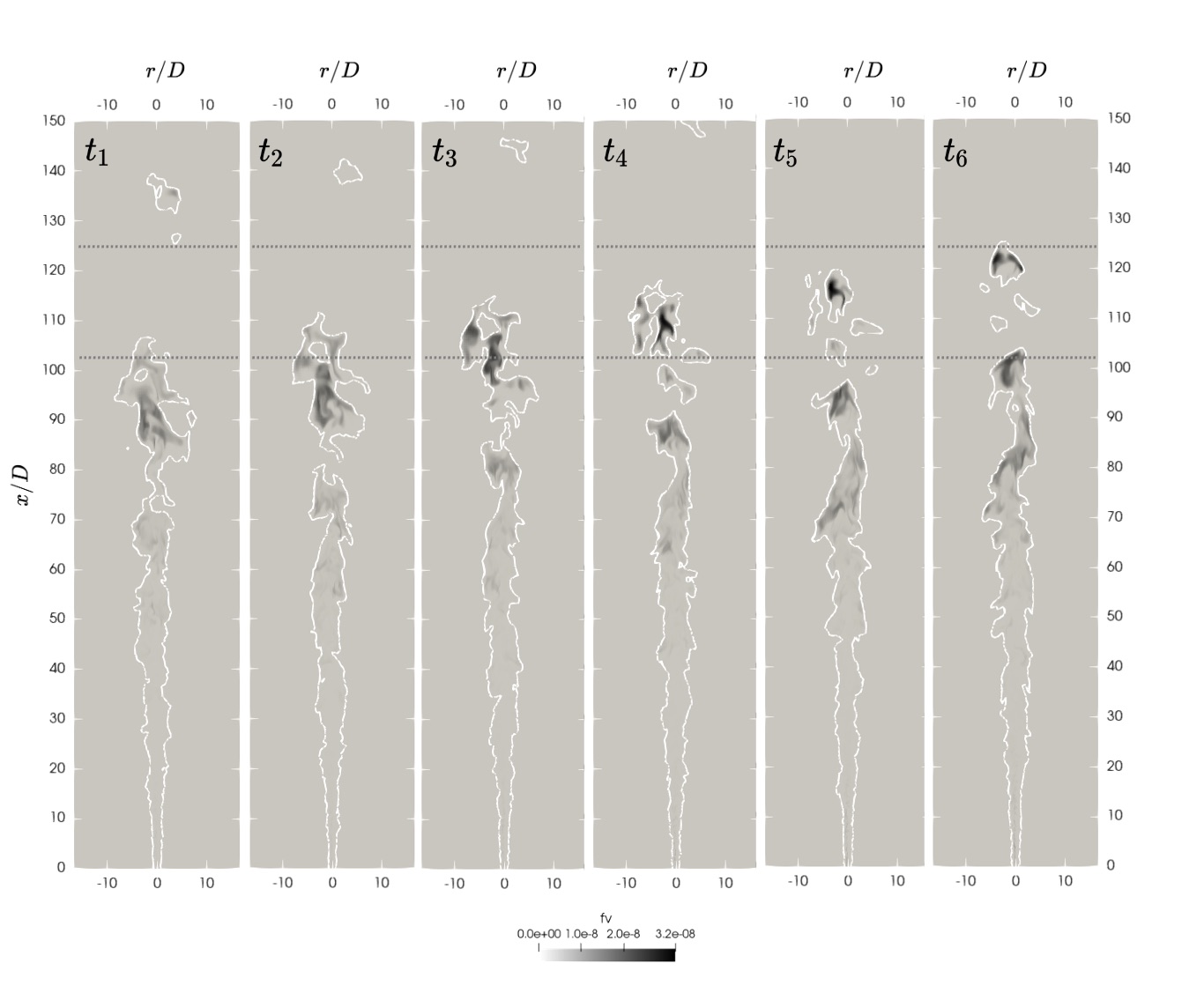}

\caption{Time evolution of the soot volume fraction. The snapshots are taken with a time interval $\Delta t=$  5.25 ms.  The white line represents the stoichiometric mixture fraction isoline  $Z=Z_{st}=$ 0.073. The dotted horizontal lines denote the axial positions $x/D=$~102 and 125, respectively, and mark the gray area from Fig.~\ref{fig:soot_Z_centerline}.}
\label{fig:TimeSeriesFv}
\end{figure}

For a qualitative assessment of the soot dynamics  downstream of $x/D=$ 102, 
instantaneous snapshots of  the soot volume fraction, taken at 5.25 ms time intervals,  are shown in Fig. ~\ref{fig:TimeSeriesFv}. It is observed that  pockets of rich mixture, enclosing soot particles,  intermittently detach from the main jet at various axial positions for $x/D>$~80. 
These   are transported downstream, where they mix with the air coflow streams, leading to the shrinking of the rich pockets and the complete  oxidation of the soot particles. 
This phenomenon repeats intermittently throughout the simulation time and seems to control the presence of the volume fraction downstream of the time-averaged stoichiometric isoline.    


\FloatBarrier 

\begin{figure}[h]
\centering
\includegraphics[width=0.6\textwidth]{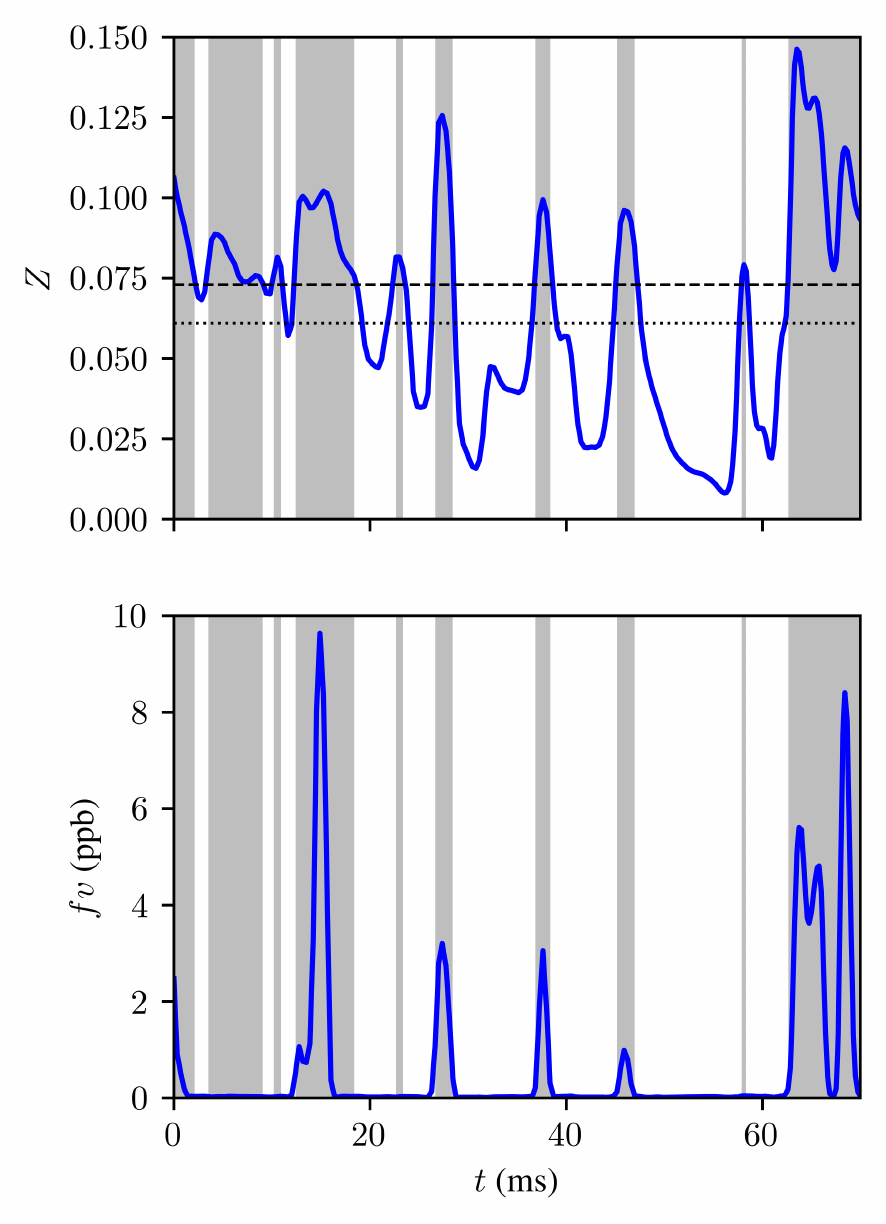}
\caption{Instantaneous values for the mixture fraction (top) and soot volume fraction (bottom) over time at an axial position $x/D=$~105. The horizontal dashed and dotted lines represent the stoichiometric mixture fraction $Z=Z_{st}=$ 0.073 and the time-averaged mixture fraction $\langle Z\rangle=$~0.061, respectively. The areas highlighted in gray correspond to rich mixture fraction samples.}
\label{fig:probes}
\end{figure}

\newpage
To further analyze this finding, mixture fraction and  soot volume fraction probes  have been collected  over 70~ms at $x/D=$~105  on the flame centerline (within the gray area of Fig.~\ref{fig:soot_Z_centerline}).
Their evolution over time  is plotted in Fig. \ref{fig:probes}, where the time instants with a mixture fraction higher than its stoichiometric value ($Z_{st}=0.073$) are highlighted in gray.  
The plots indicate strong oscillations  of the mixture fraction around its stoichiometric value, mainly correlating with the presence of a soot volume fraction above the experimental threshold. No soot volume fraction is detected outside the highlighted areas, i.e. for mixture fraction below the stoichiometric value. 

\begin{figure}[h]
\centering
\includegraphics[width=0.8\textwidth]{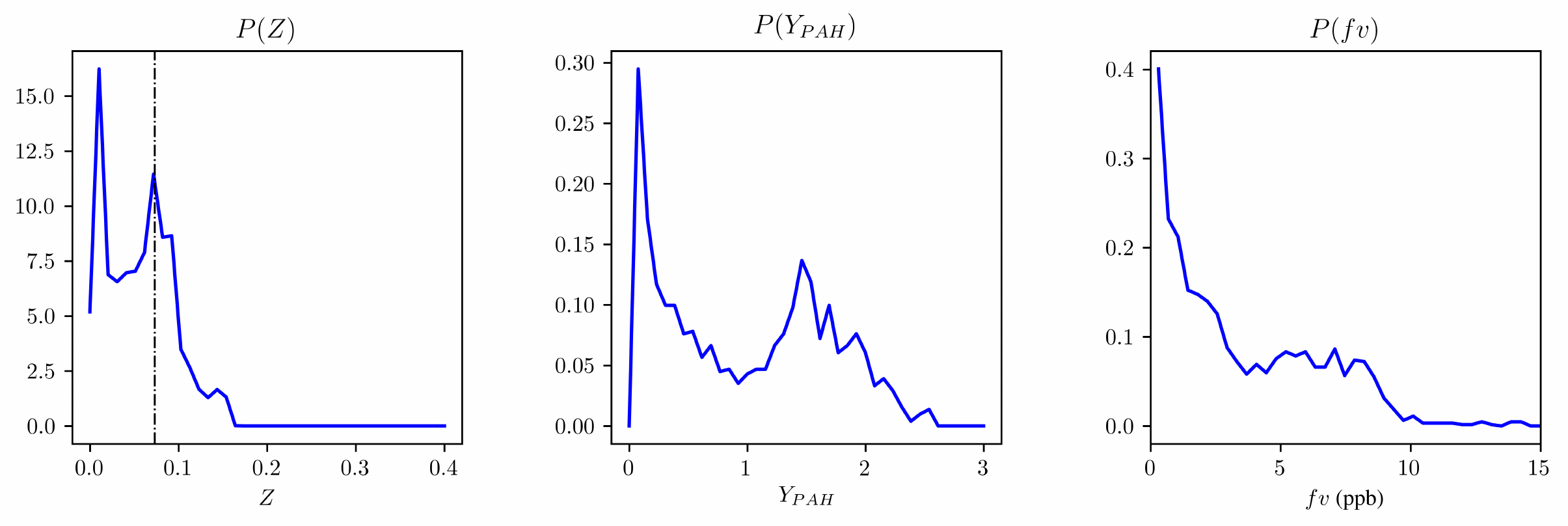}
\caption{Marginal PDF of the mixture fraction, PAH mass fraction and soot volume fraction in the neighborhood of the axial position $x/D=$ 105. The vertical dashed line in the plot of the mixture fraction PDF represents the stoichiometric mixture fraction $Z=Z_{st}=$ 0.073.}
\label{fig:PDFs}
\end{figure}

Extending the sample points in the neighborhood of the axial position $x/D=$ 105,  within 3 mm radius, makes it possible to calculate the marginal PDF of the  mixture fraction, $P(Z)$, PAH mass fraction, $P(Y_{PAH})$ and soot volume fraction, $P(f_V)$, shown in Fig.~\ref{fig:PDFs}.
 Here, all three PDFs are seen to have a bimodal  shape. In particular, the mixture fraction PDF has a first  peak at a very  lean mixture fraction value and a second, lower peak near  stoichiometry. This leads to a bimodal shape for $P(Y_{PAH})$ and for  $P(f_V)$. 
This latter result is in agreement with the experimental results presented by Qamar et al. (Fig.~8 in \cite{Qamar2009}), where the PDF of the maximum measured soot volume fraction along the centerline similarly exhibits a bimodal distribution at the flame tip. 

The above analysis explains that the experimentally observed bimodal soot volume fraction PDF, namely the  probability of finding non-sooting/sooting samples, which can be also represented as the soot intermittency,  corresponds to a bimodal probability of finding samples of lean/rich mixture fraction at that specific location in the flame. 


\FloatBarrier

\begin{figure}[h]
\centering%
\includegraphics[width=0.95\textwidth , trim=0.2cm 0.2cm 1cm 0cm,clip]{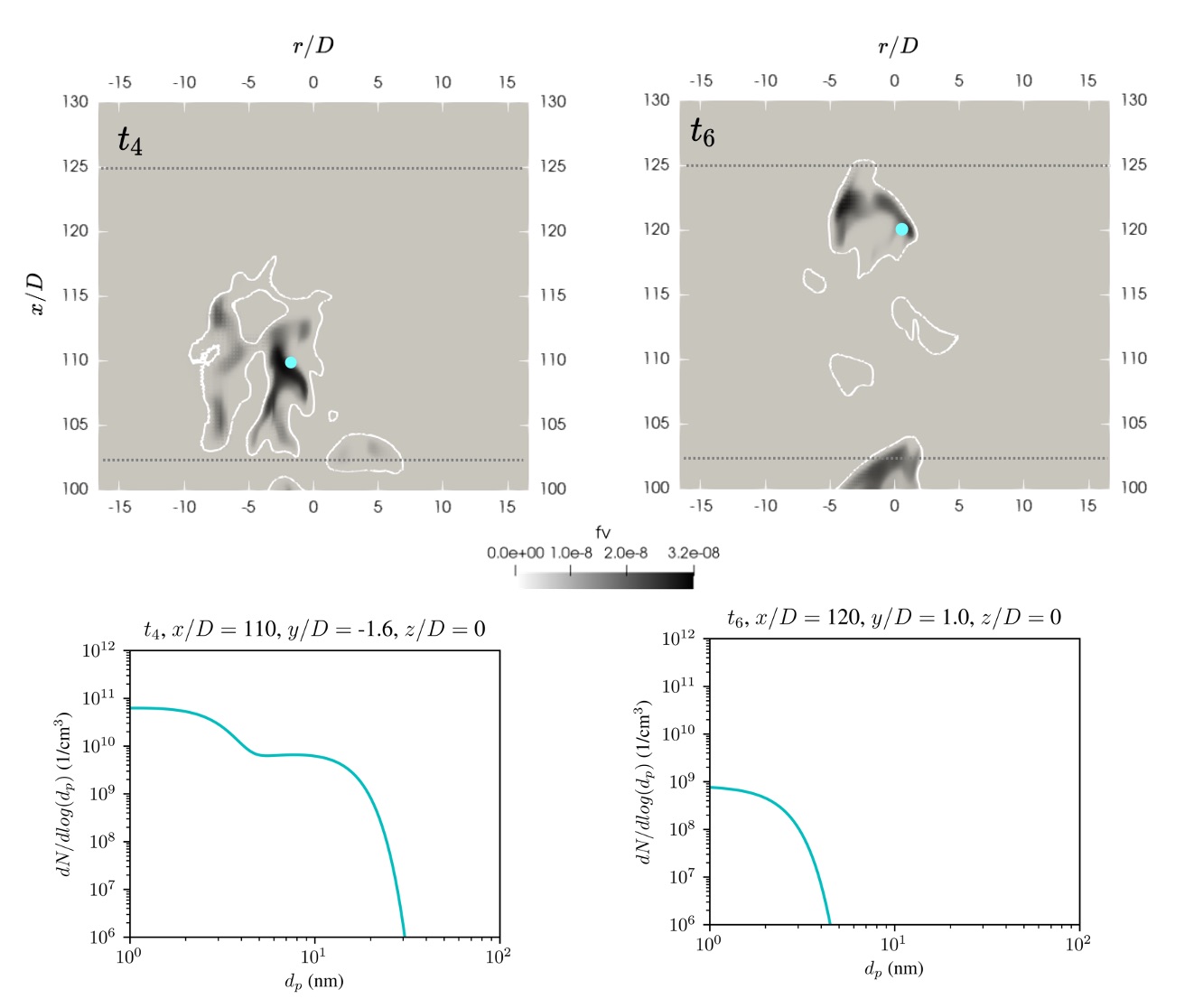}
\caption{Soot volume fraction evolution between $t_4$ (left) and  $t_6$ (right) from Fig.~\ref{fig:TimeSeriesFv} and instantaneous PSD at two arbitrarily selected points.  The white line represents the stoichiometric mixture fraction isoline  $Z=Z_{st}=$ 0.073. The dotted horizontal lines denote the axial positions $x/D=$~102 and 125, respectively.}
\label{fig:instantPSD_local}
\end{figure}
Finally, the  information on the spatial and temporal evolution of the reconstructed PSD obtained by the S-EQMOM can be further exploited to analyze the instantaneous PSD at a given location in the flame. As an example, two  points located in a detached rich mixture zone undergoing soot oxidation are arbitrarily selected at two instants in time from Fig. ~\ref{fig:TimeSeriesFv}, $t_4$ and $t_6$. The positions of these points and the  corresponding PSDs are shown in Fig.~\ref{fig:instantPSD_local}.  It is observed that  the particle number density decreases at all particle sizes between the two instants in time, due to the turbulent mixing with the lean mixture and the consequent particle oxidation. 
This example illustrates  the importance of the detailed information on the spatial and temporal PSD evolution provided by the S-EQMOM approach as a means of correctly predicting the dynamics of  soot particle formation, growth, coagulation and  oxidation. 
\FloatBarrier
\section{Conclusions}

In this study, the  Split-based EQMOM soot model was integrated into an LES solver and combined with a flamelet/progress variable (FPV) combustion model and a presumed beta-PDF for the mixture fraction.  
The main advantage of the S-EQMOM is that it provides local and continuous information on the PSD that is not accessible in standard MOM, with a  feasible computational effort. 
The LES/FPV/S-EQMOM was applied to simulate a benchmark turbulent sooting flame, the Delft Adelaide III flame. The gas phase and soot phase were  compared  with the experimental data available. Further,  a detailed analysis of the reconstructed PSD was performed at different positions along the flame.



Numerical results for the gas phase revealed very good agreement with the experimental data available in the lower portion of the flame, indicating that the flow field, the mixture and the flame structure are correctly predicted by the simulation.  
The soot volume fraction was  predicted with  good agreement,  comparable to the works of Mueller and Pitsch \cite{Mueller2012a} and Han et al.~\cite{Han2018},  and with a significant  improvement compared to other state-of-the-art approaches applied to simulate this flame. 

Predictions of the PSD along the centerline and at different radial locations indicated that there was a dominant unimodal distribution of the particle number density in the first portion of the flame. A transition from unimodal to bimodal was observed between $x/D=$ 75 and 90, close  to the position where the soot volume fraction reaches its maximum value. Nevertheless,  small particles with a  diameter lower than 100 nm  are mainly formed in this flame.  %

The detailed information from the LES was further  applied to gain insights into the soot intermittency observed in the experiments. Both the temporal evolution and the statistical behavior of the soot volume fraction with respect to the mixture fraction were investigated.  
It was found that in this slightly sooting flame, the soot intermittency observed in the experiments is mainly correlated with oscillations of the mixture fraction around its stoichiometric value close to the flame tip. Both marginal PDFs of the mixture fraction and soot volume fraction indeed exhibit a bimodal shape at this location.
 
Finally, this work demonstrates the predictive capability of the S-EQMOM in terms of both the volume fraction and the PSD of soot  in turbulent flames. The S-EQMOM therefore appears to be a promising approach to characterize the sooting features of sustainable combustion systems that have to meet limitations regarding mass and particle size distribution. 


\section*{Acknowledgments}
This research has been funded by the Clean Sky 2 Joint Undertaking under the European Union’s Horizon 2020 research and innovation programme under the ESTiMatE project, grant agreement No 821418. Calculations for this research were conducted on the Lichtenberg II Phase I high-performance computer at TU Darmstadt. 

\section*{Data Availability Statement}

The data that support the findings of this study are available from the corresponding author upon reasonable request.


\bibliographystyle{unsrtnat}
\bibliography{library_reduced}

\begin{thebibliography}{82}
\providecommand{\natexlab}[1]{#1}
\providecommand{\url}[1]{\texttt{#1}}
\expandafter\ifx\csname urlstyle\endcsname\relax
  \providecommand{\doi}[1]{doi: #1}\else
  \providecommand{\doi}{doi: \begingroup \urlstyle{rm}\Url}\fi

\bibitem[Grosschmidt et~al.(2007)Grosschmidt, Habisreuther, and
  Bockhorn]{Grosschmidt2007}
Dirk Grosschmidt, Peter Habisreuther, and Henning Bockhorn.
\newblock {Calculation of the size distribution function of soot particles in
  turbulent diffusion flames}.
\newblock \emph{Proc. Combust. Inst.}, 31 I\penalty0 (1):\penalty0 657--665,
  2007.
\newblock \doi{10.1016/j.proci.2006.07.213}.

\bibitem[Netzell et~al.(2007)Netzell, Lehtiniemi, and Mauss]{Netzell2007}
Karl Netzell, Harry Lehtiniemi, and Fabian Mauss.
\newblock {Calculating the soot particle size distribution function in
  turbulent diffusion flames using a sectional method}.
\newblock \emph{Proc. Combust. Inst.}, 31 I\penalty0 (1):\penalty0 667--674,
  jan 2007.
\newblock \doi{10.1016/j.proci.2006.08.081}.

\bibitem[Rodrigues et~al.(2018)Rodrigues, Franzelli, Vicquelin, Gicquel, and
  Darabiha]{Rodrigues2018}
Pedro Rodrigues, Benedetta Franzelli, Ronan Vicquelin, Olivier Gicquel, and
  Nasser Darabiha.
\newblock {Coupling an LES approach and a soot sectional model for the study of
  sooting turbulent non-premixed flames}.
\newblock \emph{Combust. Flame}, 190:\penalty0 477--499, 2018.
\newblock \doi{10.1016/j.combustflame.2017.12.009}.

\bibitem[Grader et~al.(2018)Grader, Eberle, Gerlinger, and Aigner]{Grader2018}
Martin Grader, Christian Eberle, Peter Gerlinger, and Manfred Aigner.
\newblock {LES of a pressurized, sooting aero-engine Model Combustor at
  different equivalence ratios with a sectional approach for PAHs and Soot}.
\newblock In \emph{ASME Turbo Expo 2018 Turbine Tech. Conf. Expo.}, pages
  GT2018--75254, 2018.

\bibitem[Tian et~al.(2021)Tian, Schiener, and Lindstedt]{Tian2021}
L.~Tian, M.A. Schiener, and R.P. Lindstedt.
\newblock {Fully coupled sectional modelling of soot particle dynamics in a
  turbulent diffusion flame}.
\newblock \emph{Proc. Combust. Inst.}, 38\penalty0 (1):\penalty0 1365--1373,
  2021.
\newblock \doi{10.1016/j.proci.2020.06.093}.

\bibitem[Cifuentes et~al.(2020)Cifuentes, Sellmann, Wlokas, and
  Kempf]{Cifuentes2020}
Luis Cifuentes, Johannes Sellmann, Iren{\"{a}}us Wlokas, and Andreas Kempf.
\newblock {Direct numerical simulations of nanoparticle formation in premixed
  and non-premixed flame-vortex interactions}.
\newblock \emph{Phys. Fluids}, 32\penalty0 (9), 2020.
\newblock \doi{10.1063/5.0020979}.

\bibitem[Zucca et~al.(2006)Zucca, Marchisio, Barresi, and Fox]{Zucca2006}
Alessandro Zucca, Daniele~L Marchisio, Antonello~A Barresi, and Rodney~O Fox.
\newblock {Implementation of the population balance equation in CFD codes for
  modelling soot formation in turbulent flames}.
\newblock 61:\penalty0 87--95, 2006.
\newblock \doi{10.1016/j.ces.2004.11.061}.

\bibitem[Attili et~al.(2014)Attili, Bisetti, Mueller, and Pitsch]{Attili2014}
Antonio Attili, Fabrizio Bisetti, Michael~E. Mueller, and Heinz Pitsch.
\newblock {Formation, growth, and transport of soot in a three-dimensional
  turbulent non-premixed jet flame}.
\newblock \emph{Combust. Flame}, 161\penalty0 (7):\penalty0 1849--1865, 2014.
\newblock \doi{10.1016/j.combustflame.2014.01.008}.

\bibitem[Mueller and Pitsch(2012)]{Mueller2012a}
Michael~E. Mueller and Heinz Pitsch.
\newblock {LES model for sooting turbulent nonpremixed flames}.
\newblock \emph{Combust. Flame}, 159\penalty0 (6):\penalty0 2166--2180, 2012.
\newblock \doi{10.1016/j.combustflame.2012.02.001}.

\bibitem[Xuan and Blanquart(2015)]{Xuan2015}
Y.~Xuan and G.~Blanquart.
\newblock {Effects of aromatic chemistry-turbulence interactions on soot
  formation in a turbulent non-premixed flame}.
\newblock \emph{Proc. Combust. Inst.}, 35\penalty0 (2):\penalty0 1911--1919,
  2015.
\newblock \doi{10.1016/j.proci.2014.06.138}.

\bibitem[Koo et~al.(2017)Koo, Hassanaly, Raman, Mueller, and Geigle]{Koo2017}
Heeseok Koo, Malik Hassanaly, Venkat Raman, Michael~E. Mueller, and Klaus~Peter
  Geigle.
\newblock {Large-Eddy Simulation of Soot Formation in a Model Gas Turbine
  Combustor}.
\newblock \emph{J. Eng. Gas Turbines Power}, 139\penalty0 (3), 2017.
\newblock \doi{10.1115/1.4034448}.

\bibitem[Sewerin and Rigopoulos(2017)]{Sewerin2017}
Fabian Sewerin and Stelios Rigopoulos.
\newblock {An LES-PBE-PDF approach for modeling particle formation in turbulent
  reacting flows}.
\newblock \emph{Phys. Fluids}, 29\penalty0 (10):\penalty0 105105, oct 2017.
\newblock \doi{10.1063/1.5001343}.

\bibitem[Sewerin and Rigopoulos(2018)]{Sewerin2018}
Fabian Sewerin and Stelios Rigopoulos.
\newblock {An LES-PBE-PDF approach for predicting the soot particle size
  distribution in turbulent flames}.
\newblock \emph{Combust. Flame}, 189:\penalty0 62--76, 2018.
\newblock \doi{10.1016/j.combustflame.2017.09.045}.

\bibitem[Seltz et~al.(2021)Seltz, Domingo, and Vervisch]{Seltz2021}
Andrea Seltz, Pascale Domingo, and Luc Vervisch.
\newblock {Solving the population balance equation for non-inertial particles
  dynamics using probability density function and neural networks: Application
  to a sooting flame}.
\newblock \emph{Phys. Fluids}, 33\penalty0 (1), 2021.
\newblock \doi{10.1063/5.0031144}.

\bibitem[Bouaniche et~al.(2019)Bouaniche, Vervisch, and
  Domingo]{Bouaniche2019a}
Alexandre Bouaniche, Luc Vervisch, and Pascale Domingo.
\newblock {A hybrid stochastic/fixed-sectional method for solving the
  population balance equation}.
\newblock \emph{Chem. Eng. Sci.}, 209:\penalty0 115198, 2019.
\newblock \doi{10.1016/j.ces.2019.115198}.

\bibitem[Han et~al.(2019)Han, Raman, Mueller, and Chen]{Han2018}
Wang Han, Venkat Raman, Michael~E. Mueller, and Zheng Chen.
\newblock {Effects of combustion models on soot formation and evolution in
  turbulent nonpremixed flames}.
\newblock \emph{Proc. Combust. Inst.}, 37\penalty0 (1):\penalty0 985--992,
  2019.
\newblock \doi{10.1016/j.proci.2018.06.096}.

\bibitem[Yoo and Im(2007)]{Yoo2007}
Chun~Sang Yoo and Hong~G. Im.
\newblock {Transient soot dynamics in turbulent nonpremixed ethylene-air
  counterflow flames}.
\newblock \emph{Proc. Combust. Inst.}, 31 I\penalty0 (1):\penalty0 701--708,
  2007.
\newblock \doi{10.1016/j.proci.2006.08.090}.

\bibitem[Lignell et~al.(2007)Lignell, Chen, Smith, Lu, and Law]{Lignell2007}
David~O. Lignell, Jacqueline~H. Chen, Philip~J. Smith, Tianfeng Lu, and
  Chung~K. Law.
\newblock {The effect of flame structure on soot formation and transport in
  turbulent nonpremixed flames using direct numerical simulation}.
\newblock \emph{Combust. Flame}, 151\penalty0 (1-2):\penalty0 2--28, oct 2007.
\newblock \doi{10.1016/j.combustflame.2007.05.013}.

\bibitem[Bisetti et~al.(2012)Bisetti, Blanquart, Mueller, and
  Pitsch]{Bisetti2012}
Fabrizio Bisetti, Guillaume Blanquart, Michael~Edward Mueller, and Heinz
  Pitsch.
\newblock {On the formation and early evolution of soot in turbulent
  nonpremixed flames}.
\newblock \emph{Combust. Flame}, 159:\penalty0 317--335, 2012.
\newblock \doi{10.1016/j.combustflame.2011.05.021}.

\bibitem[Attili et~al.(2016)Attili, Bisetti, Mueller, and Pitsch]{Attili2016}
Antonio Attili, Fabrizio Bisetti, Michael~E. Mueller, and Heinz Pitsch.
\newblock {Effects of non-unity Lewis number of gas-phase species in turbulent
  nonpremixed sooting flames}.
\newblock \emph{Combust. Flame}, 166:\penalty0 192--202, apr 2016.
\newblock \doi{10.1016/j.combustflame.2016.01.018}.

\bibitem[Raman and Fox(2016)]{Raman2016a}
Venkat Raman and Rodney~O. Fox.
\newblock {Modeling of Fine-Particle Formation in Turbulent Flames}.
\newblock \emph{Annu. Rev. Fluid Mech.}, 48\penalty0 (1):\penalty0 159--190,
  jan 2016.
\newblock \doi{10.1146/annurev-fluid-122414-034306}.

\bibitem[Valencia et~al.(2021)Valencia, Ruiz, Manrique, Celis, and {Figueira da
  Silva}]{Valencia2021}
Sebastian Valencia, Sebasti{\'{a}}n Ruiz, Javier Manrique, Cesar Celis, and
  Lu{\'{i}}s~Fernando {Figueira da Silva}.
\newblock {Soot modeling in turbulent diffusion flames: review and prospects}.
\newblock \emph{J. Brazilian Soc. Mech. Sci. Eng.}, 43\penalty0 (4):\penalty0
  1--24, 2021.
\newblock \doi{10.1007/s40430-021-02876-y}.

\bibitem[Rigopoulos(2019)]{Rigopoulos2019}
Stelios Rigopoulos.
\newblock {Modelling of Soot Aerosol Dynamics in Turbulent Flow}.
\newblock \emph{Flow, Turbul. Combust.}, 103\penalty0 (3):\penalty0 565--604,
  2019.
\newblock \doi{10.1007/s10494-019-00054-8}.

\bibitem[Mueller et~al.(2009)Mueller, Blanquart, and Pitsch]{Mueller2009a}
Michael~Edward Mueller, G~Blanquart, and H~Pitsch.
\newblock {Hybrid Method of Moments for modeling soot formation and growth}.
\newblock 156:\penalty0 1143--1155, 2009.
\newblock \doi{10.1016/j.combustflame.2009.01.025}.

\bibitem[Donde et~al.(2013)Donde, Raman, Mueller, and Pitsch]{Donde2012a}
Pratik Donde, Venkat Raman, Michael~E Mueller, and Heinz Pitsch.
\newblock {LES/PDF based modeling of soot–turbulence interactions in
  turbulent flames}.
\newblock \emph{Proc. Combust. Inst.}, 34\penalty0 (1):\penalty0 1183--1192,
  jan 2013.
\newblock \doi{10.1016/j.proci.2012.07.055}.

\bibitem[Mueller and Pitsch(2013)]{Mueller2013}
Michael~Edward Mueller and Heinz Pitsch.
\newblock {Large eddy simulation of soot evolution in an aircraft combustor
  Large eddy simulation of soot evolution in an aircraft combustor}.
\newblock \emph{Phys. Fluids}, 110812, 2013.
\newblock \doi{10.1063/1.4819347}.

\bibitem[Chong et~al.(2018)Chong, Hassanaly, Koo, Mueller, Raman, and
  Geigle]{Teng2018}
Teng~Shao Chong, Malik Hassanaly, Heeseok Koo, Michael~Edward Mueller, Venkat
  Raman, and Klaus-peter Geigle.
\newblock {Large eddy simulation of pressure and dilution-jet effects on soot
  formation in a model aircraft swirl combustor}.
\newblock \emph{Combust. Flame}, 192:\penalty0 452--472, 2018.
\newblock \doi{10.1016/j.combustflame.2018.02.021}.

\bibitem[Franzelli et~al.(2019)Franzelli, Vi{\'{e}}, and
  Darabiha]{Franzelli2018}
B~Franzelli, A~Vi{\'{e}}, and N~Darabiha.
\newblock {A three-equation model for the prediction of soot emissions in LES
  of gas turbines}.
\newblock \emph{Proc. Combust. Inst.}, 37\penalty0 (4):\penalty0 5411--5419,
  2019.
\newblock \doi{10.1016/j.proci.2018.05.061}.

\bibitem[Cokuslu et~al.(2022)Cokuslu, Hasse, Geigle, and Ferraro]{Cokuslu2022}
{\"{O}}mer~H. Cokuslu, Christian Hasse, Klaus~P. Geigle, and Federica Ferraro.
\newblock {Soot Prediction in a Model Aero-Engine Combustor using a
  Quadrature-based Method of Moments}.
\newblock In \emph{AIAA SCITECH 2022 Forum}, pages 1--12, Reston, Virginia, jan
  2022. American Institute of Aeronautics and Astronautics.
\newblock ISBN 978-1-62410-631-6.
\newblock \doi{10.2514/6.2022-1446}.

\bibitem[{Pereira Tardelli} et~al.(2021){Pereira Tardelli}, Darabiha, Veynante,
  and Franzelli]{Tardelli2021}
L{\'{i}}via {Pereira Tardelli}, Nasser Darabiha, Denis Veynante, and Benedetta
  Franzelli.
\newblock {Validating Soot Models in LES of Turbulent Flames: The Contribution
  of Soot Subgrid Intermittency Model to The Prediction of Soot Production in
  an Aero-Engine Model Combustor}.
\newblock In \emph{Vol. 3B Combust. Fuels, Emiss.}, pages 1--11. American
  Society of Mechanical Engineers, jun 2021.
\newblock ISBN 978-0-7918-8495-9.
\newblock \doi{10.1115/GT2021-60296}.

\bibitem[Eigentler et~al.(2022)Eigentler, Gerlinger, and Eggels]{Eigentler2022}
Florian Eigentler, Peter~M. Gerlinger, and Ruud Eggels.
\newblock {Soot CFD simulation of a real aero engine combustor}.
\newblock In \emph{AIAA SCITECH 2022 Forum}, Reston, Virginia, jan 2022.
  American Institute of Aeronautics and Astronautics.
\newblock ISBN 978-1-62410-631-6.
\newblock \doi{10.2514/6.2022-0489}.

\bibitem[Wick et~al.(2017{\natexlab{a}})Wick, Priesack, and Pitsch]{Wick2017b}
Achim Wick, Frederic Priesack, and Heinz Pitsch.
\newblock {Large-Eddy simulation and detailed modeling of soot evolution in a
  model aero engine combustor}.
\newblock In \emph{ASME Turbo Expo 2017}, volume 4A-2017, pages 1--10,
  2017{\natexlab{a}}.
\newblock ISBN 9780791850848.
\newblock \doi{10.1115/GT201763293}.

\bibitem[Yuan et~al.(2012)Yuan, Laurent, and Fox]{Yuan2012}
C~Yuan, F~Laurent, and R~O Fox.
\newblock {An extended quadrature method of moments for population balance
  equations}.
\newblock \emph{J. Aerosol Sci.}, 51:\penalty0 1--23, 2012.

\bibitem[Salenbauch et~al.(2015)Salenbauch, Cuoci, Frassoldati, Saggese,
  Faravelli, and Hasse]{Salenbauch2015a}
Steffen Salenbauch, Alberto Cuoci, Alessio Frassoldati, Chiara Saggese, Tiziano
  Faravelli, and Christian Hasse.
\newblock {Modeling soot formation in premixed flames using an Extended
  Conditional Quadrature Method of Moments}.
\newblock \emph{Combust. Flame}, 162\penalty0 (6):\penalty0 2529--2543, jun
  2015.
\newblock \doi{10.1016/j.combustflame.2015.03.002}.

\bibitem[Salenbauch et~al.(2016)Salenbauch, Sirignano, Marchisio, Pollack,
  Anna, and Hasse]{Salenbauch2016}
Steffen Salenbauch, Mariano Sirignano, Daniele~L Marchisio, Martin Pollack,
  Andrea~D Anna, and Christian Hasse.
\newblock {Detailed particle nucleation modeling in a sooting ethylene flame
  using a Conditional Quadrature Method of Moments ( CQMOM )}.
\newblock \emph{Proc. Combust. Inst.}, 36\penalty0 (1):\penalty0 1--9, 2016.
\newblock \doi{10.1016/j.proci.2016.08.003}.

\bibitem[Wick et~al.(2017{\natexlab{b}})Wick, Nguyen, Laurent, Fox, and
  Pitsch]{Wick2017a}
Achim Wick, Tan-trung Nguyen, Fr{\'{e}}d{\'{e}}rique Laurent, Rodney~O Fox, and
  Heinz Pitsch.
\newblock {Modeling soot oxidation with the Extended Quadrature Method of
  Moments}.
\newblock \emph{Proc. Combust. Inst.}, 36\penalty0 (1):\penalty0 789--797,
  2017{\natexlab{b}}.
\newblock \doi{10.1016/j.proci.2016.08.004}.

\bibitem[Ferraro et~al.(2021)Ferraro, Russo, Schmitz, Hasse, and
  Sirignano]{Ferraro2021a}
Federica Ferraro, Carmela Russo, Robert Schmitz, Christian Hasse, and Mariano
  Sirignano.
\newblock {Experimental and numerical study on the effect of oxymethylene
  ether-3 (OME3) on soot particle formation}.
\newblock \emph{Fuel}, 286:\penalty0 119353, feb 2021.
\newblock \doi{10.1016/j.fuel.2020.119353}.

\bibitem[Taylor and McGraw(1997)]{McGraw1997}
Publisher Taylor and Robert McGraw.
\newblock {Description of aerosol dynamics by the quadrature method of
  moments}.
\newblock \emph{Aerosol Sci. Technol.}, 27\penalty0 (2):\penalty0 255--265,
  1997.
\newblock \doi{10.1080/02786829708965471}.

\bibitem[Chalons et~al.(2010)Chalons, Fox, and Massot]{Chalons2010}
C.~Chalons, R.~O. Fox, and M.~Massot.
\newblock {A multi-Gaussian quadrature method of moments for gas-particle flows
  in a LES framework}.
\newblock \emph{Cent. Turbul. Res. Proc. Summer Progr.}, \penalty0
  (January):\penalty0 347--358, 2010.

\bibitem[Salenbauch et~al.(2019)Salenbauch, Hasse, Vanni, and
  Marchisio]{Salenbauch2019}
Steffen Salenbauch, Christian Hasse, Marco Vanni, and Daniele~L Marchisio.
\newblock {A numerically robust method of moments with number density function
  reconstruction and its application to soot formation, growth and oxidation}.
\newblock \emph{J. Aerosol Sci.}, 128:\penalty0 34--49, 2019.
\newblock \doi{10.1016/j.jaerosci.2018.11.009}.

\bibitem[Pigou et~al.(2018)Pigou, Morchain, Fede, Penet, and
  Laronze]{Pigou2018}
Maxime Pigou, J{\'{e}}r{\^{o}}me Morchain, Pascal Fede, Marie-isabelle Penet,
  and Geoffrey Laronze.
\newblock {New developments of the Extended Quadrature Method of Moments to
  solve Population Balance Equations}.
\newblock \emph{J. Comput. Phys.}, 365:\penalty0 243--268, jul 2018.
\newblock \doi{10.1016/j.jcp.2018.03.027}.

\bibitem[Nguyen et~al.(2016)Nguyen, Laurent, Fox, and Massot]{Nguyen2016}
T~T Nguyen, F~Laurent, R~O Fox, and M~Massot.
\newblock {Solution of population balance equations in applications with fine
  particles : Mathematical modeling and numerical schemes}.
\newblock \emph{J. Comput. Phys.}, 325:\penalty0 129--156, 2016.
\newblock \doi{10.1016/j.jcp.2016.08.017}.

\bibitem[Megaridis and Dobbins(1990)]{Megaridis1990a}
Constantine~M Megaridis and Richard~A Dobbins.
\newblock {A Bimodal Integral Solution of the Dynamic Equation for an Aerosol
  Undergoing Simultaneous Particle Inception and Coagulation}.
\newblock \emph{Aerosol Sci. Technol.}, 12\penalty0 (2):\penalty0 240--255, jan
  1990.
\newblock \doi{10.1080/02786829008959343}.

\bibitem[Echavarria et~al.(2011)Echavarria, Jaramillo, Sarofim, and
  Lighty]{Echavarria2011a}
Carlos~A Echavarria, Isabel~C Jaramillo, Adel~F Sarofim, and Joann~S Lighty.
\newblock {Studies of soot oxidation and fragmentation in a two-stage burner
  under fuel-lean and fuel-rich conditions}.
\newblock 33:\penalty0 659--666, 2011.
\newblock \doi{10.1016/j.proci.2010.06.149}.

\bibitem[Peeters et~al.(1994)Peeters, Stroomer, de~Vries, Roekaerts, and
  Hoogendoorn]{Peeters1994}
T.~W.J. Peeters, P.~P.J. Stroomer, J.~E. de~Vries, D.~J.E.M. Roekaerts, and
  C.~J. Hoogendoorn.
\newblock {Comparative experimental and numerical investigation of a piloted
  turbulent natural-gas diffusion flame}.
\newblock \emph{Symp. Combust.}, 25\penalty0 (1):\penalty0 1241--1248, 1994.
\newblock \doi{10.1016/S0082-0784(06)80764-2}.

\bibitem[Qamar et~al.(2009)Qamar, Alwahabi, Chan, Nathan, Roekaerts, and
  King]{Qamar2009}
N~H Qamar, Z~T Alwahabi, Q~N Chan, G~J Nathan, D~Roekaerts, and K~D King.
\newblock {Soot volume fraction in a piloted turbulent jet non-premixed flame
  of natural gas}.
\newblock \emph{Combust. Flame}, 156\penalty0 (7):\penalty0 1339--1347, 2009.
\newblock \doi{10.1016/j.combustflame.2009.02.011}.

\bibitem[Stroomer(1995)]{Stroomer1995}
P.P.J. Stroomer.
\newblock \emph{{Turbulence and OH Structures in Flames}}.
\newblock PhD thesis, Technical University Delft, 1995.

\bibitem[Marchisio and Fox(2013)]{MarchisioBook2013}
Daniele~L. Marchisio and Rodney~O. Fox.
\newblock \emph{{Computational Models for Polydisperse Particulate and
  Multiphase Systems}}.
\newblock Cambridge University Press, Cambridge, 2013.
\newblock ISBN 9781139016599.
\newblock \doi{10.1017/CBO9781139016599}.

\bibitem[Wheeler(1974)]{Wheeler1974}
John~C. Wheeler.
\newblock {Modified moments and Gaussian quadratures}.
\newblock \emph{Rocky Mt. J. Math.}, 4\penalty0 (2):\penalty0 287--296, jun
  1974.
\newblock \doi{10.1216/RMJ-1974-4-2-287}.

\bibitem[Zhao et~al.(2003)Zhao, Yang, Johnston, Wang, Anthony, Balthasar,
  Kraft, Wexler, Balthasar, and Kraft]{Zhao2003}
Bin Zhao, Zhiwei Yang, Murray~V. Johnston, Hai Wang, S~Anthony, Michael
  Balthasar, Markus Kraft, Anthony~S. Wexler, Michael Balthasar, and Markus
  Kraft.
\newblock {Measurement and numerical simulation of soot particle size
  distribution functions in a laminar premixed ethylene-oxygen-argon flame}.
\newblock \emph{Combust. Flame}, 133\penalty0 (1-2):\penalty0 173--188, 2003.
\newblock \doi{10.1016/S0010-2180(02)00574-6}.

\bibitem[Bartos et~al.(2017)Bartos, Dunn, Sirignano, D'Anna, and
  Masri]{Bartos2017}
Daniel Bartos, Matthew Dunn, Mariano Sirignano, Andrea D'Anna, and Assaad~R.
  Masri.
\newblock {Tracking the evolution of soot particles and precursors in turbulent
  flames using laser-induced emission}.
\newblock \emph{Proc. Combust. Inst.}, 36\penalty0 (2):\penalty0 1869--1876,
  2017.
\newblock \doi{10.1016/j.proci.2016.07.092}.

\bibitem[Balthasar and Kraft(2003)]{Balthasar2003}
M~Balthasar and M~Kraft.
\newblock {A stochastic approach to calculate the particle size distribution
  function of soot particles in laminar premixed flames}.
\newblock \emph{Combust. Flame}, 133\penalty0 (3):\penalty0 289--298, may 2003.
\newblock \doi{10.1016/S0010-2180(03)00003-8}.

\bibitem[Frenklach and Wang(1991)]{Frenklach1991}
Michael Frenklach and Hai Wang.
\newblock {Detailed modeling of soot particle nucleation and growth}.
\newblock \emph{Symp. Combust.}, 23\penalty0 (1):\penalty0 1559--1566, jan
  1991.
\newblock \doi{10.1016/S0082-0784(06)80426-1}.

\bibitem[Frenklach and Wang(1994)]{Frenklach1994}
Michael Frenklach and Hai Wang.
\newblock {Detailed Mechanism and Modeling of Soot Particle Formation}.
\newblock In Henning Bockhorn, editor, \emph{Soot Form. Combust.}, pages
  165--192. Springer, Berlin, Heidelberg, 1994.

\bibitem[Appel et~al.(2000)Appel, Bockhorn, and Frenklach]{Appel2000}
J{\"{o}}rg Appel, Henning Bockhorn, and Michael Frenklach.
\newblock {Kinetic modeling of soot formation with detailed chemistry and
  physics: Laminar premixed flames of C2 hydrocarbons}.
\newblock \emph{Combust. Flame}, 121\penalty0 (1-2):\penalty0 122--136, 2000.
\newblock \doi{10.1016/S0010-2180(99)00135-2}.

\bibitem[Kazakov and Frenklach(1998)]{Kazakov1998}
Andrei Kazakov and Michael Frenklach.
\newblock {Dynamic modeling of soot particle coagulation and aggregation:
  Implementation with the method of moments and application to high-pressure
  laminar premixed flames}.
\newblock \emph{Combust. Flame}, 114\penalty0 (3-4):\penalty0 484--501, 1998.
\newblock \doi{10.1016/S0010-2180(97)00322-2}.

\bibitem[Pierce and Moin(2004)]{Pierce2004}
Charles~David Pierce and Parviz Moin.
\newblock {Progress-variable approach for large-eddy simulation of non-premixed
  turbulent combustion}.
\newblock 504\penalty0 (March 2002):\penalty0 73--97, 2004.
\newblock \doi{10.1017/S0022112004008213}.

\bibitem[Ihme et~al.(2005)Ihme, Cha, and Pitsch]{Ihme2005}
Matthias Ihme, Chong~M. Cha, and Heinz Pitsch.
\newblock {Prediction of local extinction and re-ignition effects in
  non-premixed turbulent combustion using a flamelet/progress variable
  approach}.
\newblock \emph{Proc. Combust. Inst.}, 30\penalty0 (1):\penalty0 793--800, jan
  2005.
\newblock \doi{10.1016/j.proci.2004.08.260}.

\bibitem[Peters(1986)]{Peters1986}
Norbert Peters.
\newblock {Laminar Flamelet Concepts in turbulent combustion}.
\newblock \emph{Twenty-First Symp. Combust. Combust. Insititute}, pages
  1231--1250, 1986.

\bibitem[Blanquart and Pitsch(2009)]{Blanquart2009}
G~Blanquart and H~Pitsch.
\newblock {Chemical mechanism for high temperature combustion of engine
  relevant fuels with emphasis on soot precursors}.
\newblock \emph{Combust. Flame}, 156\penalty0 (3):\penalty0 588--607, 2009.
\newblock \doi{10.1016/j.combustflame.2008.12.007}.

\bibitem[Narayanaswamy et~al.(2010)Narayanaswamy, Blanquart, and
  Pitsch]{Narayanaswamy2010}
K~Narayanaswamy, G~Blanquart, and H~Pitsch.
\newblock {A consistent chemical mechanism for oxidation of substituted
  aromatic species}.
\newblock \emph{Combust. Flame}, 157\penalty0 (10):\penalty0 1879--1898, oct
  2010.
\newblock \doi{10.1016/j.combustflame.2010.07.009}.

\bibitem[Nicoud et~al.(2011)Nicoud, Toda, Cabrit, Bose, and Lee]{Nicoud2011}
Franck Nicoud, Hubert~Baya Toda, Olivier Cabrit, Sanjeeb Bose, and Jungil Lee.
\newblock {Using singular values to build a subgrid-scale model for large eddy
  simulations}.
\newblock \emph{Phys. Fluids}, 23\penalty0 (8):\penalty0 085106, aug 2011.
\newblock \doi{10.1063/1.3623274}.

\bibitem[Toda et~al.(2011)Toda, Truffin, Gilles, Cabrit, and Nicoud]{Toda2011}
Baya~Hubert Toda, Karine Truffin, Bruneaux Gilles, Olivier Cabrit, and Franck
  Nicoud.
\newblock {A dynamic procedure for advanced subgrid-scale models and
  wall-bounded flows}.
\newblock In \emph{7th Int. Symp. Turbul. Shear Flow Phenomena, TSFP 2011},
  volume 2011-July, pages 1--6, 2011.

\bibitem[Hunger et~al.(2017)Hunger, {F. Zulkifli}, {A. O. Williams}, Beyrau,
  and Hasse]{Hunger2017}
Franziska Hunger, Meor {F. Zulkifli}, Benjamin {A. O. Williams}, Frank Beyrau,
  and Christian Hasse.
\newblock {Comparative flame structure investigation of normal and inverse
  turbulent non-premixed oxy-fuel flames using experimentally recorded and
  numerically predicted Rayleigh and OH-PLIF signals}.
\newblock \emph{Proc. Combust. Inst.}, 36\penalty0 (2):\penalty0 1713--1720,
  2017.
\newblock \doi{10.1016/j.proci.2016.06.183}.

\bibitem[Popp et~al.(2015)Popp, Hunger, Hartl, Messig, Coriton, Frank, Fuest,
  and Hasse]{Popp2015}
Sebastian Popp, Franziska Hunger, Sandra Hartl, Danny Messig, Bruno Coriton,
  Jonathan~H Frank, Frederik Fuest, and Christian Hasse.
\newblock {{\{}LES{\}} flamelet-progress variable modeling and measurements of
  a turbulent partially-premixed dimethyl ether jet flame}.
\newblock \emph{Combust. Flame}, 162\penalty0 (8):\penalty0 3016--3029, aug
  2015.
\newblock \doi{10.1016/j.combustflame.2015.05.004}.

\bibitem[Gierth et~al.(2018)Gierth, Hunger, Popp, Wu, Ihme, and
  Hasse]{Gierth2018}
Sandro Gierth, Franziska Hunger, Sebastian Popp, Hao Wu, Matthias Ihme, and
  Christian Hasse.
\newblock {Assessment of differential diffusion effects in flamelet modeling of
  oxy-fuel flames}.
\newblock \emph{Combust. Flame}, 197:\penalty0 134--144, nov 2018.
\newblock \doi{10.1016/j.combustflame.2018.07.023}.

\bibitem[Wen et~al.(2021)Wen, Gierth, Rieth, Chen, and Hasse]{Wen2021d}
Xu~Wen, Sandro Gierth, Martin Rieth, Jacqueline~H. Chen, and Christian Hasse.
\newblock {Large-eddy simulation of a multi-injection flame in a diesel engine
  environment using an unsteady flamelet/progress variable approach}.
\newblock \emph{Phys. Fluids}, 33\penalty0 (10), 2021.
\newblock \doi{10.1063/5.0065351}.

\bibitem[Jones and Prasad(2010)]{Jones2010}
W.~P. Jones and V.~N. Prasad.
\newblock {Large Eddy Simulation of the Sandia Flame Series (D-F) using the
  Eulerian stochastic field method}.
\newblock \emph{Combust. Flame}, 157\penalty0 (9):\penalty0 1621--1636, 2010.
\newblock \doi{10.1016/j.combustflame.2010.05.010}.

\bibitem[Domingo et~al.(2008)Domingo, Vervisch, and Veynante]{Domingo2008}
P.~Domingo, L.~Vervisch, and D.~Veynante.
\newblock {Large-eddy simulation of a lifted methane jet flame in a vitiated
  coflow}.
\newblock \emph{Combust. Flame}, 152\penalty0 (3):\penalty0 415--432, 2008.
\newblock \doi{10.1016/j.combustflame.2007.09.002}.

\bibitem[Ihme and Pitsch(2008{\natexlab{a}})]{Ihme2008}
Matthias Ihme and Heinz Pitsch.
\newblock {Prediction of extinction and reignition in nonpremixed turbulent
  flames using a flamelet/progress variable model. 2. Application in LES of
  Sandia flames D and E}.
\newblock \emph{Combust. Flame}, 155:\penalty0 90--107, 2008{\natexlab{a}}.
\newblock \doi{10.1016/j.combustflame.2008.04.015}.

\bibitem[ISF(2021)]{ISF}
https://www.adelaide.edu.au/cet/isfworkshop/, 2021.

\bibitem[Nooren et~al.(2000)Nooren, Versluis, {Van Der Meer}, Barlow, and
  Frank]{Nooren2000}
P.~A. Nooren, M.~Versluis, T.~H. {Van Der Meer}, R.~S. Barlow, and J.~H. Frank.
\newblock {Raman-Rayleigh-LIF measurements of temperature and species
  concentrations in the Delft piloted turbulent jet diffusion flame}.
\newblock \emph{Appl. Phys. B Lasers Opt.}, 71\penalty0 (1):\penalty0 95--111,
  2000.
\newblock \doi{10.1007/s003400000278}.

\bibitem[OpenFOAM(2020)]{OpenFOAM}
OpenFOAM.
\newblock {The open source CFD toolbox, OpenFOAM}, 2020.

\bibitem[Zschutschke et~al.(2017)Zschutschke, Messig, Scholtissek, and
  Hasse]{ulf}
Axel Zschutschke, Danny Messig, Arne Scholtissek, and Christian Hasse.
\newblock {Universal Laminar Flame Solver (ULF)}.
\newblock 2017.
\newblock \doi{10.6084/m9.figshare.5119855.v2}.

\bibitem[Weise et~al.(2013)Weise, Messig, Meyer, and Hasse]{Weise2013}
Steffen Weise, Danny Messig, Bernd Meyer, and Christian Hasse.
\newblock {An abstraction layer for efficient memory management of tabulated
  chemistry and flamelet solutions}.
\newblock \emph{Combust. Theory Model.}, 17\penalty0 (3):\penalty0 411--430,
  jun 2013.
\newblock \doi{10.1080/13647830.2013.770602}.

\bibitem[Ayache and Mastorakos(2012)]{Ayache2012}
Simon Ayache and Epaminondas Mastorakos.
\newblock {Conditional Moment Closure/Large Eddy Simulation of the Delft-III
  natural gas non-premixed jet flame}.
\newblock \emph{Flow, Turbul. Combust.}, 88\penalty0 (1-2):\penalty0 207--231,
  2012.
\newblock \doi{10.1007/s10494-011-9368-6}.

\bibitem[Sweby(1984)]{Sweby1984}
P.~K. Sweby.
\newblock {High resolution schemes using flux limiters for hyperbolic
  conservation laws.}
\newblock \emph{SIAM J. Numer. Anal.}, 1984.
\newblock \doi{10.1137/0721062}.

\bibitem[Ihme and Pitsch(2008{\natexlab{b}})]{Ihme2008c}
Matthias Ihme and Heinz Pitsch.
\newblock {Modeling of radiation and nitric oxide formation in turbulent
  nonpremixed flames using a flamelet/progress variable formulation}.
\newblock \emph{Phys. Fluids}, 20\penalty0 (5), 2008{\natexlab{b}}.
\newblock \doi{10.1063/1.2911047}.

\bibitem[Schiener and Lindstedt(2018)]{Schiener2018}
Marcus~Andreas Schiener and Rune~Peter Lindstedt.
\newblock {Joint-scalar transported PDF modelling of soot in a turbulent
  non-premixed natural gas flame}.
\newblock \emph{Combust. Theory Model.}, 22\penalty0 (6):\penalty0 1134--1175,
  2018.
\newblock \doi{10.1080/13647830.2018.1472391}.

\bibitem[Huo et~al.(2022)Huo, Cleary, Masri, and Mueller]{Huo2022a}
Zhijie Huo, Matthew~J. Cleary, Assaad~R. Masri, and Michael~E. Mueller.
\newblock {A coupled MMC-LES and sectional kinetic scheme for soot formation in
  a turbulent flame}.
\newblock \emph{Combust. Flame}, 241:\penalty0 112089, 2022.
\newblock \doi{10.1016/j.combustflame.2022.112089}.

\bibitem[Ge et~al.(2013)Ge, Cleary, and Klimenko]{Ge2013}
Y.~Ge, M.~J. Cleary, and A..~Y. Klimenko.
\newblock {A comparative study of Sandia flame series (D-F) using
  sparse-Lagrangian MMC modelling}.
\newblock \emph{Proc. Combust. Inst.}, 34\penalty0 (1):\penalty0 1325--1332,
  2013.
\newblock \doi{10.1016/j.proci.2012.06.059}.

\bibitem[Raman and Pitsch(2007)]{Raman2007a}
Venkatramanan Raman and Heinz Pitsch.
\newblock {A consistent LES/filtered-density function formulation for the
  simulation of turbulent flames with detailed chemistry}.
\newblock \emph{Proc. Combust. Inst.}, 31\penalty0 (2):\penalty0 1711--1719,
  jan 2007.
\newblock \doi{10.1016/j.proci.2006.07.152}.

\end{thebibliography}

\end{document}